\documentclass[usenatbib]{mn2e} 
\pdfoutput=1
\usepackage{graphicx}
\usepackage[paperwidth=8.5in, paperheight=11in,top=1in, bottom=1in, left=1in, right=1in]{geometry}
\usepackage[authoryear]{natbib}

\bibpunct{(}{)}{;}{a}{}{;} 

\usepackage[T1]{fontenc}

\usepackage{amsfonts}
\usepackage{amsmath}
\usepackage{txfonts}

\usepackage{fleqn} 

\usepackage{booktabs}

\title{Radio Source Evolution on Galactic Scales}
\author[T. Maciel and P. Alexander]{T. Maciel$^{1}$\thanks{E-mail:
tm419@cam.ac.uk (TM)} and P.
Alexander$^{1}$\\
$^{1}$Astrophysics Group, Cavendish Laboratory, J. J. Thomson Ave, Cambridge CB3 0HE, UK}

\begin{document}

\date{Accepted 2014 June 3. Received 2014 May 7; in original form 2014 March 14} 
\pagerange{\pageref{firstpage}--\pageref{lastpage}} \pubyear{2014}
\maketitle
\label{firstpage}


\begin{abstract}
There is mounting evidence that mechanical radio source feedback is important in galaxy evolution and in order to quantify this feedback, detailed models of radio source evolution are required. We present an extension to current analytic models that encompasses young radio sources with physical sizes on sub-kiloparsec scales. This work builds on an existing young source dynamical model to include radiative losses in a flat environment, and as such, is the best physically-motivated Compact Symmetric Object model to date. Results predict that young radio sources experience significant radiative loss on length scales and spectral scales consistent with observed Compact Steep-Spectrum sources. We include full expressions for the transition to self-similar expansion and present this complete model of radio source evolution from first cocoon formation to end of source lifetime around $10^{8}$ years within the context of a simplified King profile external atmosphere. 

\end{abstract}




\section{Introduction}\label{s:intro}
Understanding the detailed evolution of radio-loud jets and cocoons as they propagate through the intragalactic and intracluster medium (IGM/ICM) is vital in order to quantify the importance of mechanical radio source feedback in galaxy and cluster evolution. It is clear that expanding jets and cocoons do interact with their environment, hollowing out huge cavities in the X-ray-emitting intracluster medium, sending weak shock waves out to larger radii \citep{Fabian_2006}, and inducing turbulence in the cloudy environment \citep{Krause_2007,Antonuccio-Delogu_2008}. For reviews, see \citet{Fabian_2012} and \citet{McNamara_2012}. Estimates of the kinetic energy available in these extended radio sources match or exceed the cooling rates expected from the hot IGM/ICM gas (\citealt{Rafferty_2006}; \citealt{McNamara_2007}; \citealt*{Rafferty_2008}) and thus mechanical feedback is a strong candidate to offset the `cooling flow problem' in clusters \citep{Fabian_1994} and reduce star formation rates in the galaxy (\citealt{Silk_1998}; \citealt*{Quilis_2001}), regulating overall galaxy growth through a feedback loop (\citealt{Best_2006}; \citealt{Croton_2006}; \citealt*{Shabala_2011}). On large intergalactic scales successful analytic models exist which determine the lifetime, growth, and energetics of classical double radio sources. But to address kinetic-mode feedback on galactic scales, a new model of young source evolution is required to keep pace with the growing numbers of sub-kiloparsec scale observations and simulations of AGN feedback (e.g. \citealt{Phillips_1980}; \citealt{Phillips_1982}; \citealt{Wilkinson_1994}; \citealt*{Taylor_1996}; \citealt{Gugliucci_2005}; \citealt{Guo_2011}; \citealt{Kunert-Bajraszewska_2010}; \citealt{An_2012a}; \citealt{Ciotti_2012}; \citealt*{Wagner_2012}; \citealt{Dallacasa_2013}).

The structures of double radio sources vary greatly in size and are observed from intragalactic parsec scales out to the intracluster environment megaparsecs away from the AGN core. For scales upwards of several kiloparsecs, the radio source morphology is strikingly similar such that it is impossible to determine the true physical size or age based purely on the morphology (see studies of source size with axial ratio by \citealt{Leahy_1984} and \citealt*{Leahy_1989}). This similarity suggests a self-similar solution \citep{Falle_1991} and forms the basis of existing analytic models of classical double sources. Falle shows that if the external environment density $\rho_{\rmn{x}}$ decreases slower than $D^{-2}$ (which is the rate of density decrease for a ballistic jet of size $D$), then eventually $\rho_{\rmn{j}}<\rho_{\rmn{x}}$ and the environment acts to reconfine the jet. 

\citet[hereafter KA97]{Kaiser_1997} extend this model to show that the pressure in the building cocoon, rather than the external medium, can reconfine the jet in a feedback process that is essential to self-similar evolution. The dynamical model of KA97 is further developed to include radiative processes in \citet*[hereafter KDA97]{Kaiser_1997b}. This model takes as primary input the radio source environment and initial jet power (assumed constant) and allows for loss mechanisms in the form of synchrotron ageing, adiabatic expansion of the cocoon, and inverse-Compton scattering off CMB photons. Evolved over the lifetime of a source, the model produces the source size and radio luminosity as a function of age, as well as detailed evolution of the cocoon dynamics. Varying the initial conditions generates separate evolutionary tracks -- a process which can be used to simulate entire populations of double radio sources.

Variations on the self-similar KDA97 model which examine in greater detail the dynamics of the hotspot and the evolution of the injected electron spectra have subsequently been presented by \citealt{Komissarov_1998}; \citealt*{Blundell_1999}; \citealt{Manolakou_2002}; and \citealt{Barai_2007}. Observational tests of the leading evolutionary models find good overall agreement with the KDA97 model. A study by \citet{Barai_2006} which compares simulated radio source surveys to the complete samples of 3CRR, 6CE, and 7CRS finds that data simulated using the KDA97 model gives a better fit to observed surveys than the \citet{Blundell_1999} or \citet{Manolakou_2002} models, although \citet{Barai_2007} propose a modification to the KDA97 model that allows for the evolution of the hotspot radius with time in order to improve the fit. Recently \citet*{Kapinska_2012} have used a similar approach to determine the most likely intrinsic parameter values of powerful radio sources based on goodness-of-fit with observed populations. 

These large-scale analytic models provide a crucial link to observed radio source dynamics and spectra, but how do the young radio sources evolve on galactic scales less than a few kpc? Answering this question is crucial for quantifying the mechanical AGN feedback within a host galaxy and is the aim of this paper. Observationally, there are four commonly-used terms to classify young radio sources, reviewed in \citet{Augusto_2006}. Radio morphologies from VLBI suggest that Compact Symmetric Objects (CSOs) exist on scales $\la$ 1~kpc \citep{Wilkinson_1994,Readhead_1996} and Medium Symmetric Objects (MSOs) on scales between 1~kpc and $\sim$ 15~kpc \citep{Fanti_1995,Augusto_2006}. These form the precursors to the more mature large-scale FRI and FRII sources \citep{Fanaroff_1974}. From a spectral view, Gigahertz Peaked Spectrum (GPS) sources with sizes < 1~kpc (\citealt*{ODea_1991}; \citealt{ODea_1998}) have a peak around 1 GHz, while Compact Steep Spectrum (CSS) sources have sizes between 1 and $\sim$ 15~kpc and a spectral peak closer to 100 MHz \citep{Fanti_1995}. There is large overlap between these classifications in the sense that most CSOs are also GPS sources (\citealt*{Bicknell_1997}; \citealt{Ostorero_2010}) and the MSO class contains the larger GPS objects and CSS objects. It is usually proposed that CSOs (GPS objects) evolve into MSOs (CSS objects), with the most powerful eventually becoming the classical FRIIs at large scales and the rest either shutting off completely or evolving into lower-luminosity, turbulent FRIs \citep{Fanti_1995,Readhead_1996,Alexander_2000,Snellen_2000,Kunert-Bajraszewska_2010a,An_2012}. We will refer to these classes when considering spectral predictions based on our new young source model.

Early work modelling the emission of young radio galaxies assumed self-similar expansion in a homogeneous, falling atmosphere ($r^{-\beta}$ where $\beta$ ranges between 1 and 2) and predicted a decreasing luminosity as the young source expanded \citep{Begelman_1996}. But current galaxy density profiles from X-ray studies suggest a profile consistent with a King profile \citep{King_1962} where the central density is roughly constant out to a few kiloparsecs \citep{Diaz_2005}. Realistically, the parsec scale environment is not homogeneous but rather found to contain colder gas clumps within a hot interstellar medium \citep{Guillard_2012, Tacconi_2013}, but for the purpose of radio source analytic models, this inhomogeneity is usually ignored. However some hydrodynamical simulations of jets in a non-uniform medium have been performed \citep{Wiita_2004, Sutherland_2007, Krause_2007,Wagner_2011}.

A dense core environment coincident with the sub-kiloparsec scales of Compact Symmetric Objects (CSOs) will greatly affect how the source expands. A flat environment is considered in \citet{Kaiser_2007b}, and more recently by \citet{An_2012}, assuming self-similar expansion. Using the KDA97 radiative model applied to both flat and falling environments, both studies predict strong synchrotron losses for young radio sources and provide instructive commentary for the young source regime. But the dynamics still assume a self-similar evolution which we expect to break down on sub-kiloparsec scales. This is because a characteristic length scale relevant to the sizes of young radio sources can be defined by the primary input parameters (\citealt{Falle_1991} and KA97). Thus self-similar evolution is not expected for young sources and a more complex model is necessary. 

A semi-analytic dynamical model for young source evolution which departs from self-similarity was developed by \citet[hereafter A06]{Alexander_2006} and tracks the evolution from the point of cocoon formation up until jet reconfinement and the subsequent transition to self-similar expansion. Here, we extend this dynamical model to include radiative evolution using the methodology of KDA97 to track energy losses within the cocoon. We find that radiative losses are significant for young radio sources within a constant density environment.

A strong tool for visualizing radio source evolution is the radio power versus source size ($P{-}D$) diagram, introduced by \citet{Shklovskii_1963,Baldwin_1982}. It tracks two easily-derived parameters (the radio luminosity and projected source size), tracing the evolution of a radio source over the course of its lifetime in a method analogous to the Hertzsprung--Russell diagram of stellar evolution. This is the tool we shall use to visualize our analytic evolutionary models.  

This paper is structured as follows. The young model dynamics of A06 are reviewed in section \ref{s:dynamical model} along with the transition to a self-similar evolution in flat and falling density environments. Section \ref{s:radiative model} presents the radiative model extension to young radio sources, starting with the assumption of rough equipartition within the cocoon and allowing for radiative losses in the older parts of the cocoon. The results are presented and discussed in sections \ref{s:results} and \ref{s:discussion}, and a summary is given in section \ref{s:summary}. We assume a power-law synchrotron spectra in the extended radio structure following $F_{\rmn{\nu}}\propto \nu^{-\alpha}$, where $F_{\rmn{\nu}}$ is the radiative flux and $\alpha$ is the spectral index.
%
%
%
\section{Young Source Evolution}\label{s:dynamical model}
We consider the situation where a ballistic jet emerges from the central engine, expanding at relativistic speeds into an isotropic environment with a density gradient following a simplified King profile with a core density $\rho_{\rmn{0}}$
\begin{equation}\label{e:environment}
\rho_{\rmn{x}} =
\left\{
  \begin{array}{ll}
    \rho_{\rmn{0}} & \rmn{if}\, r \leq a_{\rmn{0}} \\
    \rho_{\rmn{0}}\left(\frac{a_{\rmn{0}}}{r}\right)^{\beta} & \rmn{if}\, r > a_{\rmn{0}}
  \end{array}
\right.
\end{equation}
such that $\rho_{\rmn{0}}\,a_{\rmn{0}}^{\beta}$ becomes a characteristic quantity
of the external medium. Here $a_{\rmn{0}}$ is the core radius of the host
galaxy, typically around a few kiloparsecs in agreement with X-ray measures of galactic density profiles (\citealt*{Canizares_1987}; \citealt*{Fukazawa_2004}; \citealt*{Kawakatu_2008}) and studies of hotspot size with distance \citep*{Kawakatu_2009a}. 

\begin{figure*}
\centering{}
\includegraphics[width=0.99\textwidth]{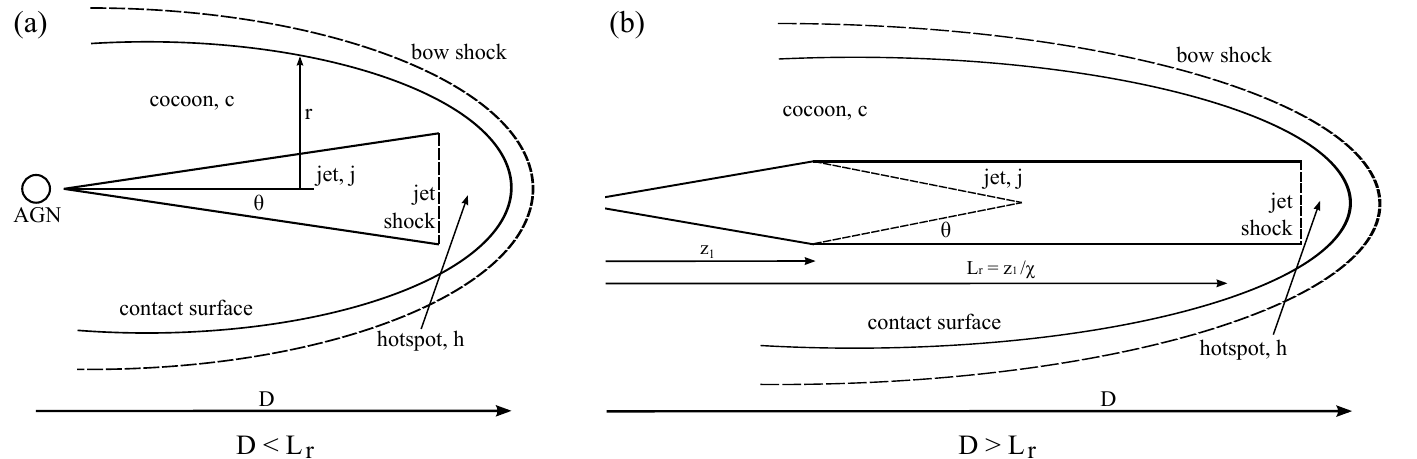}
\caption{The model geometry and notation used in this paper. Figure (a) is a snapshot schematic of a young radio source with an unconfined, conical jet and surrounding cocoon. Dynamics follow the model presented in section \ref{s:dynamics}. Figure (b) is a snapshot diagram of a much older radio source, after the jet is fully reconfined by the cocoon pressure and evolution begins to follow the self-similar model reviewed in section \ref{s:transition}. Diagrams not to scale.}
\label{f:model schematic}    
\end{figure*}

The end of the jet shocks the IGM/ICM at the hotspot, and once the jet material becomes under-dense, a backflow of material at the hotspot inflates enormous cocoons filled with a plasma of particles and magnetic fields. The jet pushes forward and the cocoon expands outwards as the source ages so that the overall size increases with time. A schematic diagram for young radio sources is shown in figure \ref{f:model schematic}, along with the length scales we discuss in section \ref{s:length scales}. Throughout this paper the following notation is used: subscripts j, h, c, and x refer to the relevant property of the jet, hotspot, cocoon, and external environment, respectively.

  \subsection{Length scales}\label{s:length scales}
  The basic ingredients for a young radio source are assumed to be the same as for a larger self-similar source. The jet is characterized by its constant kinetic power $Q\,=\,\frac{1}{2}\dot{M}\varv_{\rmn{j}}^{2}$, opening angle $\theta$, and mass transport rate, $\dot{M}$. The source expansion is governed by the external environment profile, $\rho_{\rmn{x}}$. This paper is concerned with young sources still within their galactic core so here we assume a flat density profile, $\rho_{\rmn{x}}\,=\,\rho_{\rmn{0}}$. From the above parameters we can write down a characteristic length scale (cf. Falle 1991; KA97; A06):
  \begin{equation}\label{e:L1}
  L_{1}\propto\left(\frac{\dot{M}^{3}}{\rho_{0}^{2}\,Q}\right)^{1/4} \equiv 2\sqrt{2}\left(\frac{Q}{\rho_{0}\,\varv_{\rmn{j}}^{3}}\right)^{1/2}
  \end{equation}
  where the right-hand side comes from substituting $\dot{M}=2Q/\varv_{\rmn{j}}^{2}$ and following A06, we choose the $2\sqrt{2}$ factor for later convenience. Typical values for jet power ($Q = 1.3\times10^{39}$~W), speed ($\varv_{\rmn{j}} = c$), and external density ($\rho_{\rmn{0}} = 7.2\times10^{-22}$) give $L_{1}\,=\,24$~pc. For previously-developed self-similar models, this length is tiny compared to the relevant scales of classical double sources (kiloparsec to megaparsec scales) but here we seek to develop a young source model on galactic scales and we can no longer ignore this characteristic size. Thus we do not expect self-similar evolution for young sources. As shown in A06 (equation 4), physically this length scale is similar to the scale when the jet density begins to fall below the constant external density, occurring at $L_{\rmn{c}}=L_{1}/2\Omega^{1/2}$ where $\Omega\sim \piup \theta^{2}$ is the solid angle of the conical jet cross-section.

  Beyond $L_{\rmn{c}}$ the jet material is under-dense and a cocoon begins to inflate from the backflow of material at the hotspot. This onset of cocoon formation is where we begin our young source model. The source will evolve until the pressure in the cocoon is high enough compared to the falling sideways ram pressure of the jet as to send an oblique reconfinement shock to the jet axis \citep{Falle_1991}. KA97 show that this self-confinement process between the cocoon and the jet is crucial for self-similar evolution, and subsequent expansion should no longer depend on the initial mass transport along the jet. The process of reconfinement depends upon the dynamics of the young jet and cocoon, so we discuss the relevant reconfinement length scale, $L_{\rmn{r}}$, in more detail after reviewing the A06 dynamical model in section \ref{s:dynamics}.

  Following full reconfinement, we end our young source model and join it to the self-similar model of KA97. This transition is discussed in section \ref{s:transition}. The now self-similarly evolving source will continue to expand and radiate according to the KDA97 model until its end of life at much larger scales. Observational constraints on the maximum source age come from both dynamical and spectral ageing measurements and are in rough agreement, implying a typical FRII end of life around $t_{\rmn{end}}\,=\,10^{8}$ years (\citealt{Blundell_1999}; \citealt*{Bird_2008}; \citealt{ODea_2009a}; \citealt*{Antognini_2012}), at which point we assume the jet switches off.

  \subsection{Young source dynamics}\label{s:dynamics}
  The semi-analytic equations governing the dynamics of the young source from first cocoon formation to jet reconfinement are given in A06 and rederived here for clarity.
  In this young source regime, the external environmental density is constant, $\rho_{\rmn{x}}=\rho_{\rmn{0}}$.

  To plot any track on the $P{-}D$ diagram we need, at a minimum, expressions for how the source size, pressure, and volume evolve with age. From these three, assumption of equipartition inside the cocoon can give a minimum energy value for radio luminosity and thus a track through the $P{-}D$ diagram when coupled with size evolution (see section \ref{s:min energy}). In section \ref{s:energy losses} we refine this simple track to include radiative losses throughout the cocoon.

  By considering the shock conditions of the hotspot and bow shock at the end of the jet, the densities and velocities on either side of the shocks can be related (cf. A06 section 3.1). A differential equation for length can be solved to give an expression for young source size evolution with source age (cf. A06 equation 8)
  \begin{equation}
  D_{1}\left(t\right)=\frac{L_{1}}{K_{1}}\left(\sqrt{\frac{2K_{1}\varv_{\rmn{j}}t}{L_{1}}+1}-1\right)\label{e:young size}
  \end{equation}
  where $K_{1}$ is a constant given by A06 equation (10).
  The length evolution explicitly depends on the characteristic length scale $L_{1}$. For $D_{1} \gg L_{1}$, the size expands approximately as $t^{1/2}$ similar to \citet{Scheuer_1974}.

  The evolution of the cocoon pressure, $p_{\rmn{c}}$, and volume, $V_{\rmn{c}}$, come from considering the energy of the radio source, where the input energy from the jet, $Q\,\rmn{d}t$, is offset by the change in internal energy of the cocoon, $\rmn{d}(u_{\rmn{int}}V_{\rmn{c}})$, and the work done by the expanding cocoon on the environment, where $u_{\rmn{int}}$ is the internal energy density of the cocoon.
  \begin{equation}Q\,\rmn{d}t=\rmn{d}\left(u_{\rmn{int}}V_{\rmn{c}}\right) + p_{\rmn{c}}\rmn{d}V_{\rmn{c}} \label{e:energy equation}
  \end{equation}
  A06 neglects the energy of the hotspots for reasons argued in their section 3: the energy stored in the hotspot can be shown to increase less quickly than the total energy injected, $Q\,t$, for $D_{1} \gg L_{1}$ and in fact the energy stored in the cocoon is significantly greater than the energy in the hotspot for all $D_{1} > L_{1}$. We explore a simple model for hotspot luminosity evolution in section \ref{s:hotspot model} and show that hotspot contribution is only significant on scales less than or near $L_{\rmn{c}}$. Therefore we neglect it here. Future models of young `naked' jets before cocoon formation cannot ignore the dominant hotspot contribution however.

  Equation (\ref{e:energy equation}) can be rewritten assuming the cocoon material undergoes ideal, adiabatic expansion so that $u_{\rmn{int}}=p_{\rmn{c}}/\left(\Gamma_{\rmn{c}}-1\right)$. Then
  \begin{equation}Q=\frac{1}{\Gamma_{\rmn{c}}-1} V_{\rmn{c}} \frac{\rmn{d}p_{\rmn{c}}}{\rmn{d}t}+\frac{\Gamma_{\rmn{c}}}{\Gamma_{\rmn{c}}-1}p_{\rmn{c}}\frac{\rmn{d}V_{\rmn{c}}}{\rmn{d}t}\label{e:energy equation 2}
  \end{equation}  
  Sounds speeds within the cocoon are assumed to be large enough such that the pressure at a given time is uniform throughout the cocoon, as discussed in KA97 and A06.
  An additional important assumption, first made by \citet{Scheuer_1974}, is that the width of the shocked gas is small so that the contact surface is close to the bow shock and the radial expansion of the cocoon, $\dot{r}$, is governed by ram pressure confinement $p_{\rmn{c}} \approx \rho_{\rmn{x}}\dot{r}^{2}$. 
  Then the radius of the cocoon at distance $z$ from the core follows A06 equation (13)
  \begin{equation}
  r=\intop_{t(D=z)}^{t} \left(\frac{p_{\rmn{c}}}{\rho_{\rmn{x}}}\right)^{1/2}\rmn{d}t \label{e:radius}
  \end{equation}
  If the cocoon volume is cylindrical then
  \begin{equation}
  V_{\rmn{c}}=\intop_{L_{\rmn{c}}}^{D} \piup r^{2}\,\rmn{d}z \label{e:volume w radius}
  \end{equation}
  where $z$ is along the length of the cocoon.

  A06 determines the cocoon pressure and volume in the regimes of $D_{1} \ll L_{1}$ and $D_{1} \gg L_{1}$ using equations (\ref{e:energy equation 2}), (\ref{e:radius}), and (\ref{e:volume w radius}), and with these limiting expressions finds an approximate solution by requiring the pressure and volume evolution to be smooth and continuous throughout the young source model (see A06 equations 26 to 27 and the discussion thereafter). Thus the young source cocoon pressure evolves as
  \begin{equation}
  p_{\rmn{c}1}\left(t\right)=a_{2}\,\Omega^{-1/4}\frac{Q}{\varv_{\rmn{j}}L_{1}^{2}}\left[\frac{D_{1}\left(t\right)}{L_{1}}\right]^{-1}\left[\lambda+\frac{D_{1}\left(t\right)}{L_{1}}\right]^{-1/2} \label{e:young pressure}
  \end{equation}
  where $\lambda = \left(a_{2}/a_{4}\right)^{2}\Omega^{-1/2}$ comes from matching the form of $p_{\rmn{c}1}$ to the limiting cases of very small and very large source sizes described in A06, and $a_{2}$ and $a_{4}$ are the constants of proportionality give by A06 equations (21) and (25).
  
  The young source volume evolves as 
  \begin{equation}V_{\rmn{c}1}\left(t\right)= a_{1}\,\Omega^{3/4}L_{1}^{3}\left[\frac{D_{1}\left(t\right)}{L_{1}}\right]^{2}\left[\omega+\frac{D_{1}\left(t\right)}{L_{1}}\right]^{3/2}\label{e:young volume} 
  \end{equation} 
  where $\omega = \left(a_{3}/a_{1}\right)^{2/3}\Omega^{-1/2}$ comes from matching the form of $V_{\rmn{c}}$ to the limiting cases (note that A06 uses the term $\beta$ instead of $\omega$). The constants $a_{1}$ and $a_{3}$ are given by A06 equations (20) and (24).
  Evolution of the young radio source continues until the cocoon pressure fully reconfines the jet. Once the reconfinement process is complete, the jet pressure equals the cocoon pressure, the density of the collimated jet is constant along its length for a particular time, and the forward momentum of the source has increased relative to the unconfined young jet. The source expands self-similarly beyond $L_{\mathrm{r}}$.

  \subsection{Transition to self-similar evolution}\label{s:transition}
  The task is now to link the dynamical young source model of A06 to the self-similar expansion presented in KA97 and KDA97.

  Physically, we expect the source age, size, cocoon pressure, and cocoon volume to be smooth and continuous functions across the boundary between the young source and the self-similar regimes, with some complex changes in dynamics during the reconfinement process. However in this model, we are matching two different analytic models and expect a discontinuity at the boundary, $L_{\rmn{r}}$. While a full model for reconfinement is beyond the scope of this paper, \citet{Komissarov_1998} work out in detail the dynamics of the reconfinement shock. We use the more approximate argument derived in A06 (see also \citealt{Falle_1991} and KA97), and discussed below.

  A06 determine the start of jet reconfinement, $z_{1}=\chi L_{\rmn{r}}$, as the length scale where the cocoon pressure equals the sideways ram pressure of the jet, $p_{\rmn{c}}=p_{\rmn{j}}$ (here we use $\chi$ instead of the $\alpha$ used in A06). $L_{\rmn{r}}$ is the source size at the time, $t_{\rmn{r}}$, when the jet is fully reconfined and $\chi$ is the parameter roughly determining how many shock structures must fit inside the jet before the jet is fully reconfined (see figure \ref{f:model schematic}). A06 adopt $\chi\sim 1/3$ so that full reconfinement does not occur until the source size is greater than one and a half shock structures of length $2 z_{1}$. To derive $L_{\rmn{r}}$, A06 equate the sideways ram pressure of the unconfined jet at $z_{1}$ (A06 equation 29) with the young source cocoon pressure (equation \ref{e:young pressure}) and solve the subsequent quadratic equation for $L_{\rmn{r}}/L_{1}$.
  \begin{equation}L_{\rmn{r}}\sim \frac{8\Omega^{1/2}}{\piup^{2} a_{2}^{2} \chi ^{4}(\Gamma_{\rmn{j}}+1)^{2}}  \left(1+\sqrt{1+\frac{\piup^{2}a_{2}^{4}\chi^{4}(\Gamma_{\rmn{j}}+1)^{2}}{4 a_{4}^{2}\Omega}}\right)L_{1} \label{e:reconfinement}
  \end{equation}
  A06 equation (30) is equivalent to this $L_{\rmn{r}}$ expression (excepting a minus sign typo that is subsequently amended), if $\Gamma_{\rmn{j}}=4/3$ is substituted in. Parameters $a_{2}$ and $a_{4}$ relate to the evolution of the young source cocoon pressure derived in A06 and discussed in the next section. For typical values (see table \ref{t:parameters}) and assuming $\chi = 1/3$, full reconfinement occurs on scales of several hundred parsecs. 

  For the purposes of analytically linking young source evolution to the self-similar model of KA97, we simplistically assume that the young source model we develop here is valid up until full reconfinement, $L_{\rmn{r}}$, at which point we transition the source to a self-similar solution. This simplification allows us to explore the full lifetime of a radio source from cocoon formation out to large-scale self-similar evolution. 

  The constants determining self-similar evolution will be fixed by the conditions at full jet reconfinement, $L_{\rmn{r}}$, as we match age, size, pressure, and volume across the boundary. Whether this boundary occurs within a flat environment ($L_{\rmn{r}} < a_{\rmn{0}}$) or falling environment depends upon the length of the reconfinement process, and the specific value of $a_{\rmn{0}}$ for a given radio source. Large values of $Q$ and $\theta$, or low values of $\rho_{0}$, can push $L_{\rmn{r}}$ out to scales typical of a galaxy core radius (as discussed in section \ref{s:discussion}). Otherwise, the radio source will first reconfine at $L_{\rmn{r}}$ within the flat density of the galaxy core and follow self-similar evolution for $\beta=0$. When the source reaches $a_{\rmn{0}}$, a second transition to a falling external density ($\beta > 0$) will occur. \citet{Alexander_2000} showed that self-similar evolution will be maintained throughout this environment transition. 

  Thus we assume that three phases for full radio source evolution exist: (1) young source evolution from $L_{\rmn{c}}$ to $L_{\rmn{r}}$, (2) self-similar evolution in a flat density profile from $L_{\rmn{r}}$ to $a_{\rmn{0}}$, and (3) self-similar evolution in a falling density profile from $a_{\rmn{0}}$ to the end of source lifetime and a maximum size, $L_{\rmn{end}}$, of several megaparsecs. We outline the transitions between phases below. 

  \textbf{Transition from (1) to (2):} Self-similar expansion is governed by A06 equation (31), similar to KA97 equation (4) but with the constants derived from the boundary conditions at $L_{\rmn{r}}$ and setting $\beta=0$. The naming convention of A06 is kept for the constants whenever possible. Phase 2 source size evolves as
  \begin{equation}
  D_{2}\left(t\right)=c_{1}L_{1}\left(\frac{t-t_{1}}{\tau_{1}}\right)^{3/5}\label{e:flat self-similar size}
  \end{equation}
  where $\tau_{1} \equiv L_{1}/\varv_{\rmn{j}}$ and the $t_{1}$ term is included to match source size at the boundary, such that $t$ still represents the true age in the self-similar regime. $L_{1}$ is the natural length scale to use here instead of the $a_{\rmn{0}}$ used in KA97 since the source has not yet reached the core radius. 
  Requiring equations (\ref{e:young size}) and (\ref{e:flat self-similar size}) to match at time $t_{\rmn{r}}$ when $D_{1}\left(t_{\rmn{r}}\right)\,=\,L_{\rmn{r}}$ we solve for $t_{1}$
  \begin{equation}
  t_{1}=t_{\rmn{r}}-\tau_{1}\left[\frac{1}{K_{1}c_{1}}\left(\sqrt{\frac{2K_{1}\varv_{\rmn{j}}\,t_{\rmn{r}}}{L_{1}}+1}-1\right)\right]^{5/3} \label{e:t1}
  \end{equation}
  where $c_{1}$ comes from matching the pressure across the boundary.
  \begin{figure}
  \centering{}
  \includegraphics[width=0.9\columnwidth]{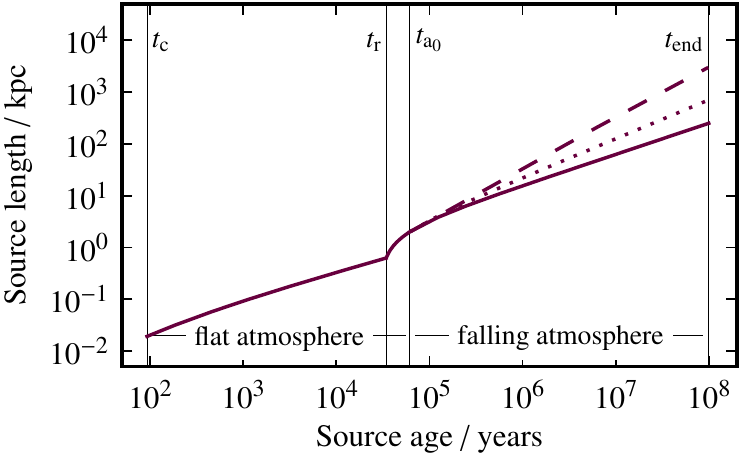}
  \caption{Radio source size evolution across the three evolutionary phases (equation \ref{e:full length}), matching source size at the boundaries with a continuous source age, $t$. Vertical lines mark the age boundaries between the three phases discussed in section \ref{s:transition}. In the third phase, the external density gradient is varied: the dashed line has $\beta$=1.9; the dotted line has $\beta$=1.0; and the solid line maintains the flat density profile ($\beta$=0) of the previous phases. The other parameters used are: $Q=1.3\times10^{39}$~W, $\theta=20^{\rmn{\circ}}$, and as given in table \ref{t:parameters}.}
  \label{f:size plot}    
  \end{figure}
  From KA97, cocoon pressure equals the post reconfinement jet pressure and evolves with source size as (cf. KA97 equations 4 and 20)
  \begin{equation}
  p_{\rmn{c}2}\left(t\right)=\frac{18}{25\left(\Gamma_{\rmn{x}}+1\right)}\frac{\left(c_{1}L_{1}\right)^{10/3}}{c_{2}} \rho_{\rmn{0}} \left(\frac{\theta}{\tau_{1}}\right)^{2} D_{2}\left(t\right)^{-4/3} \label{e:flat self-similar pressure with length}
  \end{equation}
  The constant $c_{1}$ is determined at jet reconfinement, such that the cocoon pressure is piecewise continuous across $L_{\rmn{r}}$.
  \begin{equation}
  c_{1}=\frac{1}{L_{1}}\left[c_{\rmn{p}}L_{\rmn{r}}^{4/3} \left(\frac{L_{\rmn{r}}}{L_{1}}\right)^{-1} \left(\lambda+\frac{L_{\rmn{r}}}{L_{1}}\right)^{-1/2}\right]^{3/10} \label{e:c1}
  \end{equation}
  where
  \begin{equation}
  c_{\rmn{p}}=\frac{25\, a_{2}\,\Omega^{-1/4} c_{2}\tau_{1}^{2}\,Q \left(\Gamma_{\rmn{x}}+1\right)}{18\, \theta^{2} \rho_{\rmn{0}}\, \varv_{\rmn{j}}\,L_{1}^{2}}
  \end{equation}
  Determined by the shock conditions at the hotspot, $c_{2}$ is given by KA97 equation (17).
  \begin{figure*}
  \centering{}
  \includegraphics[width=0.8\textwidth]{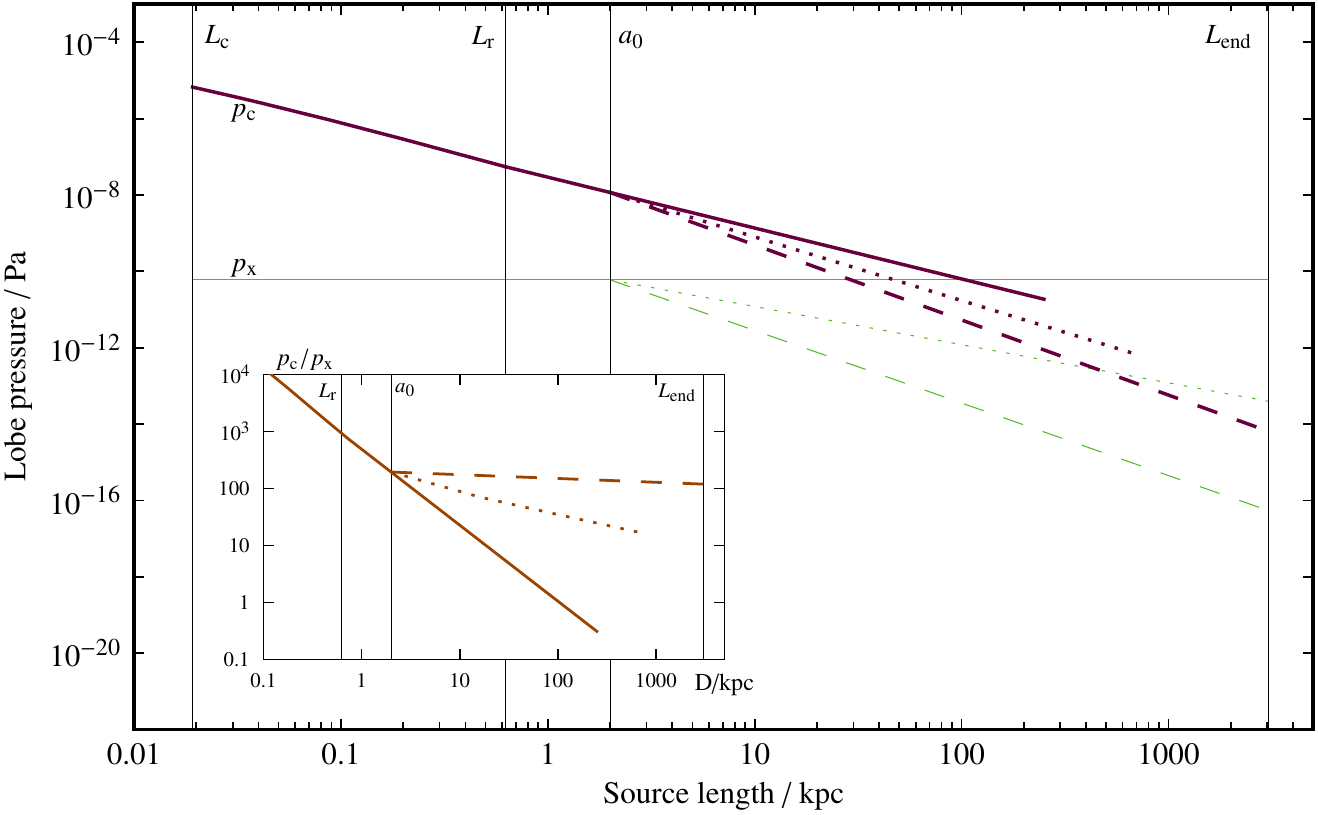}
  \caption{Cocoon pressure evolution across the three evolutionary phases (equation \ref{e:full pressure}), matching the pressure at the boundaries. For comparison the external pressure profile (thin, green lines), assuming an isothermal atmosphere, follows the same form as the simplified King profile for external density: flat in the core and falling as $r^{-\beta}$ after the transition to self-similar evolution. Vertical lines mark the size boundaries between the three phases discussed in section \ref{s:transition}. In the third phase, the external density gradient is varied: the dashed line has $\beta=1.9$; the dotted line has $\beta=1.0$; and the solid line maintains the flat density profile ($\beta=0$) from the previous phase. Cocoon pressure tracks are evaluated until $t_{\rmn{end}}=10^{8}$ years. The inset plot shows the cocoon overpressure ratio, $p_{\rmn{c}}/p_{\rmn{x}}$, as a function of source size, assuming a flat core for $D<a_{0}$ and a density gradient of $\beta$ for $D\ge a_{0}$. In the third phase, the dashed line has $\beta=1.9$, the dotted line has $\beta=1.0$, and the solid line has $\beta=0$.
  For all tracks we use: $Q=1.3\times10^{39}$~W, $\theta=20^{\rmn{\circ}}$, $p_{\rmn{0}}=6\times10^{-11}$~Pa, and other parameters as given in table \ref{t:parameters}.}
  \label{f:pressure plot}    
  \end{figure*}
  The cocoon volume in the self-similar regime follows (cf. KA97 equation 22)
  \begin{equation}
  V_{\rmn{c}2}\left(t\right)=c_{3}D_{2}\left(t\right)^{3}\label{e:flat self-similar volume}
  \end{equation}
  where $c_{3}$ sets the geometry of the cocoon and is generally less than the spherical case of $4\piup/3$. Here $c_{3}$ is determined from the volume at reconfinement, $D_{1}\left(t_{\rmn{r}}\right)\,=\,L_{\rmn{r}}$, matching equation (\ref{e:young volume}) and equation (\ref{e:flat self-similar volume})
  \begin{equation}
  c_{3}=a_{1}\,\Omega^{3/4}\frac{L_{1}}{L_{\rmn{r}}}\left(\omega + \frac{L_{\rmn{r}}}{L_{1}}\right)^{3/2} \label{e:c3}
  \end{equation}
  Constants $t_{1}$, $c_{1}$, and $c_{3}$ match the source size, pressure, and volume across the $L_{\rmn{r}}$ boundary while the source age $t$ is continuous. Equations (\ref{e:flat self-similar size}) through (\ref{e:c3}) describe the self-similar source evolution in a flat environment, with constants determined by the conditions at jet reconfinement.

  \textbf{Transition from (2) to (3):} A self-similar source in a flat density environment evolving beyond $a_{\rmn{0}}$ into a falling environment (with $\beta\,>\,0$) undergoes a transition that we treat in a completely analogous way as discussed above -- we match source size, pressure, and volume across $a_{\rmn{0}}$ with new normalizing constants now determined by the conditions at time $t_{\rmn{a_{0}}}$ when $D_{2}\left(t_{\rmn{a_{0}}}\right) = a_{\rmn{0}}$.
  \begin{equation}
  D_{3}\left(t\right)=b_{1}L_{1}\left(\frac{t-t_{2}}{\tau_{1}}\right)^{3/(5-\beta)}\label{e:falling self-similar size}
  \end{equation}
  where $t_{2}$ matches the source size at the boundary
  \begin{equation}
  t_{2}=t_{\rmn{a}_{\rmn{0}}}-\tau_{1}\left(\frac{c_{1}}{b_{1}}\right)^{(5-\beta)/3}\,\left(\frac{t_{\rmn{a}_{\rmn{0}}}-t_{1}}{\tau_{1}}\right)^{(5-\beta)/5} \label{e:t2}
  \end{equation}
  The cocoon pressure continues to evolve self-similarly in phase 3, but now in a falling atmosphere parametrized by $\beta$:
  \begin{equation}
  p_{\rmn{c}3}\left(t\right)=\frac{18 \left(b_{1}L_{1}\right)^{(10-2\beta)/3}}{\left(\Gamma_{\rmn{x}}+1\right)\left(5-\beta\right)^{2} c_{2}} \rho_{\rmn{0}} a_{\rmn{0}}^{\beta} \left(\frac{\theta}{\tau_{1}}\right)^{2} D_{3}\left(t\right)^{-(4+\beta)/3} \label{e:falling self-similar pressure with length}
  \end{equation}
  The constant $b_{1}$ is determined from the conditions at $a_{\rmn{0}}$, such that the cocoon pressure is piecewise continuous across $a_{\rmn{0}}$
  \begin{equation}
  b_{1}=\frac{1}{L_{1}}\left(\frac{(5-\beta)^{2}}{25} (c_{1} L_{1})^{10/3} a_{\rmn{0}}^{-(2\beta)/3}\right)^{3/(10-2\beta)} \label{e:e1}
  \end{equation}
  The cocoon volume in phase 3 evolves self-similarly with $D_{3}(t)$
  \begin{equation}
  V_{\rmn{c}3}\left(t\right)=b_{3}D_{3}\left(t\right)^{3}\label{e:falling self-similar volume}
  \end{equation}
  where $b_{3}$ = $c_{3}$ to match the volume across $a_{\rmn{0}}$.
  Equations (\ref{e:falling self-similar size}) through (\ref{e:falling self-similar volume}) describe the self-similar source evolution in a falling environment, with constants determined by the conditions at the core radius, $a_{\rmn{0}}$.

\begin{figure}
  \centering{}
  \includegraphics[width=0.9\columnwidth]{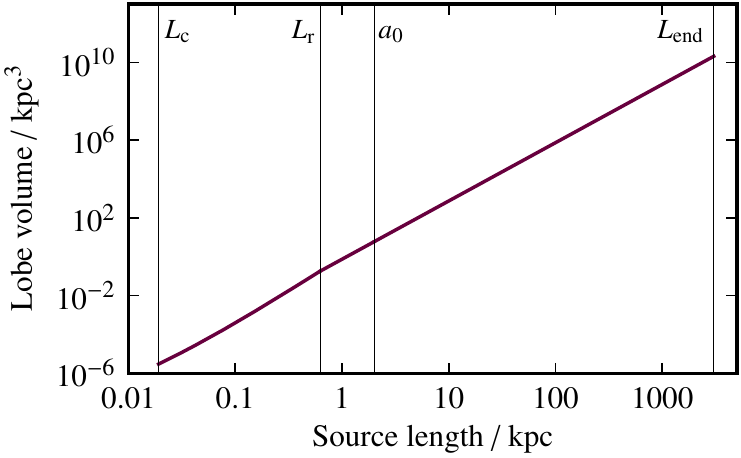}
  \caption{Cocoon volume evolution with source size is shown by the solid blue line for all three evolutionary phases (equation \ref{e:full volume}), matching the total volume at the boundaries. In the first and second phase $\beta$=0, and in the third phase $\beta$=1.9. The other parameters used are: $Q=1.3\times10^{39}$~W, $\theta=20^{\rmn{\circ}}$, and as given in table \ref{t:parameters}.}
  \label{f:volume plot}    
  \end{figure}

  Thus the full expressions of length, pressure, and volume are:
  \begin{equation}\label{e:full length}
  D\left(t\right) =
  \left\{
    \begin{array}{ll}
      D_{1}\left(t\right)& \rmn{if}\, t_{\rmn{c}} \leq t \leq t_{\rmn{r}}\\
      D_{2}\left(t\right)& \rmn{if}\, t_{\rmn{r}} < t \leq t_{\rmn{a}_{\rmn{0}}}\\
      D_{3}\left(t\right)& \rmn{if}\, t > t_{\rmn{a}_{\rmn{0}}}
    \end{array}
  \right.
  \end{equation}
  \begin{equation}\label{e:full pressure}
  p_{\rmn{c}}\left(t\right) =
  \left\{
    \begin{array}{ll}
      p_{\rmn{c}1}\left(t\right)& \rmn{if}\, t_{\rmn{c}} \leq t \leq t_{\rmn{r}}\\
      p_{\rmn{c}2}\left(t\right)& \rmn{if}\, t_{\rmn{r}} < t \leq t_{\rmn{a}_{\rmn{0}}}\\
      p_{\rmn{c}3}\left(t\right)& \rmn{if}\, t > t_{\rmn{a}_{\rmn{0}}}
    \end{array}
  \right.
  \end{equation}
  \begin{equation}\label{e:full volume}
  V_{\rmn{c}}\left(t\right) =
  \left\{
    \begin{array}{ll}
      V_{\rmn{c}1}\left(t\right)& \rmn{if}\, t_{\rmn{c}} \leq t \leq t_{\rmn{r}}\\
      V_{\rmn{c}2}\left(t\right)& \rmn{if}\, t_{\rmn{r}} < t \leq t_{\rmn{a}_{\rmn{0}}}\\
      V_{\rmn{c}3}\left(t\right)& \rmn{if}\, t > t_{\rmn{a}_{\rmn{0}}}
    \end{array}
  \right.
  \end{equation}
  These piecewise continuous expressions for radio source size, pressure, and volume are shown in figures \ref{f:size plot}, \ref{f:pressure plot}, and \ref{f:volume plot}, respectively.

  An important assumption of this model is that the cocoon pressure remains greater than the external pressure, $p_{\rmn{x}}$, for the duration of the source lifetime. In figure \ref{f:pressure plot} we plot a typical estimate of the external pressure profile (thin, green lines), assuming an isothermal atmosphere, which follows the same form as the simplified King profile for external density: flat in the core and falling as $p_{0}(a_{0}/r)^{\beta}$ after the transition to self-similar evolution. In figure \ref{f:pressure plot}, $\beta$ is varied between 0 and 1.9. The core pressure in a hydrostatic external medium is determined by the core density (here we use $\rho_{\rmn{0}}=7\times10^{-22}$ kg m$^{-3}$) and temperature ($T_{\rmn{0}}=10^{7}$~K) such that $p_{\rmn{0}}=(k_{B} T_{\rmn{0}} \rho_{\rmn{0}})/\overline{m} = 6\times10^{-11}$~Pa, where $\overline{m}$ is the mass of the proton. With these values, we plot $p_\rmn{c}/p_\rmn{x}$ in figure \ref{f:pressure plot} inset, and find that this pressure ratio is much greater than one for the duration of the source lifetime for all but the flattest of external pressure gradients (i.e. $\beta\,<\,1$). 

  The choice of $T_{0}=10^{7}$~K follows the usual assumption made by other radio source analytic papers \citep{Blundell_1999,Kaiser_2007b,Luo_2010}, but is likely an upper limit at sub-kiloparsec scales (i.e. hydrodynamical simulations of galactic environments by \citet{Barai_2013} find galaxy profile temperatures ranging between $10^{5}$ and $10^{7}$~K). A lower core temperature only makes it even more unlikely that the cocoon pressure drops below the external pressure during a typical radio source lifetime. For simplicity, we assume an isothermal atmosphere (i.e. \citealt{Giacintucci_2008}). Cosmological simulations incorporating radio source expansion during and after the active AGN phase by \citet{Barai_2008} also find that the cocoon remains over-pressurized with respect to the external medium, even long after AGN activity ceases. Their external pressure comes from N-body simulations of baryonic gas which follows the dark matter in a $\mathrm{\lambdaup}$CDM universe. The reader is referred to \citep{Eilek_1989,Alexander_2002,Kaiser_2007b} for studies of cocoon evolution when $p_\rmn{c} \sim p_\rmn{x}$. 
  %

\section{Radiative Evolution}\label{s:radiative model}

The radiation from the jets and cocoons of radio galaxies is dominated by synchrotron emission. The usual relations and assumptions are outlined below. See
Chapter 8 of \citet{Longair_2011} for a full derivation. 
We begin with the synchrotron energy loss rate of a high energy electron in a magnetic field of energy density $u_{\rmn{B}}$, averaged over all pitch angles from $\left(0,\,\piup\right)$
\begin{equation}
-\left(\frac{\rmn{d}E}{\rmn{d}t}\right)=\frac{4}{3}\sigma_{\rmn{T}} c u_{\rmn{B}} \left(\frac{v_{\rmn{e}}}{c}\right)^{2}\gamma^{2}
\end{equation}
where $\sigma_{\rmn{T}}$ is the Thomson cross-section, $v_{\rmn{e}}$ is the speed of the electron, and $\gamma$ is the associated Lorentz factor such that the energy of the electron is $E=\gamma m_{\rmn{e}} c^{2}$.

Over a power-law distribution of electron energies,
\begin{equation}\label{e:number distribution}
n(\gamma)\rmn{d}\gamma=n_{\rmn{0}}\gamma^{-p}\rmn{d}\gamma
\end{equation}
where $n(\gamma)\rmn{d}\gamma$ is the number density of electrons in energy interval
$\rmn{d}\gamma$, we make the usual approximation that electrons radiate only at the critical synchrotron frequency, 
\begin{equation}\label{e:critical frequency}
\nu \approx \nu_{\rmn{c}} \approx \gamma^{2}\nu_{\rmn{g}}=\gamma^{2}\left(\frac{eB}{2\piup m_{\rmn{e}}}\right)
\end{equation}
where $\nu_{\rmn{g}}$ is the gyrofrequency. With this approximation, the emissivity per unit volume over a distribution of electron energies is $J(\rmn{\nu})\,\rmn{d}\nu=(-\rmn{d}E/\rmn{d}t)\,n(\gamma)\,\rmn{d}\gamma$ and the total power radiated per frequency per solid angle is (cf. KDA97 equation 2)
\begin{equation}\label{e:dPower}
P_{\rmn{\nu}}=\frac{J(\rmn{\nu}) V_{\rmn{c}}}{4\piup} = \frac{\sigma_{\rmn{T}}\, c\, u_{\rmn{B}}}{6\piup \nu} \gamma^{3}n\left(\gamma \right)V_{\rmn{c}}
\end{equation}
where we have assumed $v_{\rmn{e}}=c$. This has the usual proportionality that $P_{\rmn{\nu}}~\propto~\nu^{-(p-1)/2}~\propto~\nu^{-\alpha}$ where $\alpha$ is the spectral index.

  \subsection{Loss-less minimum energy tracks}\label{s:min energy}
  We want to express the radiation in terms of the most important source
  parameters: size, pressure, and volume. The usual approach when modelling the expansion of a volume of synchrotron-emitting electrons is to assume the co-evolution of the particle and magnetic field energy densities such that $u_{\rmn{B}} = 3/4\,u_{\rmn{particle}}$. This corresponds to the minimum energy requirement derived in chapter 16 of \citet{Longair_2011}. Starting with equation (\ref{e:dPower}), it can be
  shown that the minimum energy for a volume of radiating electrons is
  \begin{equation}
  E_{\rmn{total}}\left(\rmn{min}\right)=\frac{7}{6\mu_{\rmn{0}}}V^{3/7}\left[\frac{3\mu_{\rmn{0}}}{2}G(p)\eta\, P_{\rmn{\nu}}\right]^{4/7}\label{e:Etotal}
  \end{equation}
  $\mu_{\rmn{0}}$ is the vacuum permeability and $\eta=1+\beta_{\rmn{e}}$, where Longair has defined $\beta_{\rmn{e}}$ as the ratio of the proton to electron energy densities. $G(p)$ is a constant that depends on $p$, $\nu_{\rmn{min}}$, and $\nu_{\rmn{max}}$
  \begin{equation}\label{e:G_p}
  G(p)=\left(\frac{2\piup m_{\rmn{e}}}{e}\right)^{1/2}\left(\frac{6 \mu_{\rmn{0}} c e}{(2-p)\sigma_{\rmn{T}}}\right)\left(\nu_{\rmn{max}}^{(2-p)/2}-\nu_{\rmn{min}}^{(2-p)/2}\right) \nu^{(p-1)/2}
  \end{equation}

  Expressing the energy in terms of energy density, $u_{\rmn{min}}=E_{\rmn{tot}}\left(\rmn{min}\right)/V_{\rmn{c}}$, allows
  the introduction of cocoon pressure. Assuming an ideal, adiabatic fluid in the cocoon, then
  \begin{equation}
  u_{\rmn{min}}=\frac{p_{\rmn{c}}}{\left(\Gamma_{\rmn{c}}-1\right)}\label{e:Umin pressure}
  \end{equation}
  Combining (\ref{e:Etotal}) with (\ref{e:Umin pressure}) and solving
  for $P_{\rmn{\nu}}$ gives the typical minimum energy result that the radio power per unit frequency and solid angle is proportional to $V_{\rmn{c}}\, p_{\rmn{c}}^{7/4}$
  \begin{equation}
  P_{\rmn{\nu}}(\rmn{min})=\frac{\frac{2}{3}\left(\frac{6}{7}\right)^{7/4}\mu_{\rmn{0}}^{3/4}}{G(p)\,\eta} V_{\rmn{c}}\left(\frac{p_{\rmn{c}}}{\Gamma_{\rmn{c}}-1}\right)^{7/4}\label{e:min energy luminosity}
  \end{equation}
  Using the piecewise continuous expressions for total pressure and volume (equations \ref{e:full pressure} \& \ref{e:full volume}), we can express the source luminosity as function of jet length across the full radio source lifetime in this simplest radiative model, ignoring energy losses. This `loss-less' track through the $P{-}D$ diagram is shown as the dotted lines in figure \ref{f:PD plot} for $Q\,=\,1.3\times10^{39}$~W, $\beta\,=\,0$ for the young model regime, and $\beta\,=\,1.9$ for self-similar evolution, presented for a range of observing frequencies, $\nu=\nu_{\rmn{obs}}$. Other initial parameters are given in tables \ref{t:parameters} and \ref{t:derived parameters}. 

  KDA97 explored the effect that heavier particles can have on the $P{-}D$ tracks of radio sources by varying $\beta_{\rmn{e}}$ (they use parameter name $k'$) between 0 and 100 in their figure 7. The addition of protons leads to a strong decrease in luminosity and, as KDA97 discuss, would require a radio source to have an unusually high jet power in order to match the highest observed luminosities. This also implies a much larger source size than typically observed and as a result, KDA97 suggest protons in jets are unlikely. Hence we adopt a negligible contribution from heavy particles so that $u_{\rmn{particle}} \sim u_{\rmn{e}}$ and $\eta=1$.

  KDA97 conclude in their section 3 that the equation of state of the cocoon (including the radiating particles, the thermal particles, and the magnetic fields) does not significantly impact the evolutionary track through the $P{-}D$ diagram. Here we assume the relativistic equation of state ($\Gamma_{\rmn{c}}=\Gamma_{\rmn{B}}=4/3$) which, certainly for the young source model will be realistic. For simplicity, we also use this equation of state in the self-similar regime, although it is possible that the true equation of state varies with source age and position within the cocoon.

  Of course equation (\ref{e:min energy luminosity}) is a simplistic model and
  adiabatic expansion and radiative losses are also significant. We incorporate these in the following section.

  \subsection{Radiative losses}\label{s:energy losses}
  Following the approach of KDA97, we now introduce energy losses to this minimum energy model and sum over small volume packets.
  Relativistic electrons at a particular energy $E=\gamma m_{\rmn{e}}c^{2}$ will radiate their energy away over a finite lifetime after which they no longer contribute to the total emission. Thus a distribution of electron energies will show signs of radiative loss over time. The geometry of the radio source means radiative losses will be particularly significant at the back of the cocoon for older electrons while the head of the cocoon is continually injected with a power-law distribution of fresh electrons following equation (\ref{e:number distribution}).
  Thus the time of injection, $t_{\rmn{i}}$ becomes important, as well as the source age, $t$. With this in mind we rewrite equation (\ref{e:dPower}):
  \begin{equation}
  P_{\rmn{\nu}}=\frac{\sigma_{\rmn{T}}\, c\, u_{\rmn{B}}}{6\piup\nu}\gamma^{3}n\left(t,\, t_{\rmn{i}}\right)V_{\rmn{c}}\left(t,\, t_{\rmn{i}}\right)\label{e:dPower ti}
  \end{equation}
  Relevant to young source scales there are two dominant loss mechanisms which change the electron energy with time: adiabatic expansion of the cocoon which decreases the strength of the flux-frozen magnetic field in the cocoon and synchrotron ageing from the energy lost through radiation over time. For the young source model, inverse-Compton (IC) scattering losses are negligible since the energy density of the electrons in small cocoons and at local redshifts is much higher than the CMB photon energy density ($u_{\rmn{CMB}}=\left(4\sigma_{\rmn{T}} T_{\rmn{CMB}}^{4}\right)/c\,\times \left(1+z\right)^{4}$ where $T_{\rmn{CMB}}\,\approx\,2.725$~K). In the later self-similar stages, IC losses become important when $u_{\rmn{e}}\sim u_{\rmn{CMB}}$, leading to the sharp `knee' break in the radio luminosity at large source sizes. We include IC effects here for continuity when transitioning to the later phases.

  In analogy to KDA equation (4), the energy losses are expressed through the rate of change of the electron Lorentz factor with time.
  \begin{equation}
  \frac{\rmn{d}\gamma}{\rmn{d}t}=\frac{\gamma}{3 \Gamma_{\rmn{c}}}\frac{1}{p_{\rmn{c}}\left(t\right)}\frac{\rmn{d}p_{\rmn{c}}\left(t\right)}{\rmn{d}t}-\frac{4\sigma_{\rmn{T}}}{3 m_{\rmn{e}}c}\gamma^{2}\left(u_{\rmn{B}}+u_{\rmn{CMB}}\right)\label{e:gamma evolution}
  \end{equation}
  The right hand side includes adiabatic losses from a relativistic particle in an expanding volume, $V_{\rmn{c}}\propto p_{\rmn{c}}(t)^{-1/\Gamma_{\rmn{c}}}$ \citep[chapter 16]{Longair_2011}, as well as radiative synchrotron and IC losses which have the usual dependency on the magnetic field energy density, $u_{\rmn{B}}$, and CMB photon energy density, $u_{\rmn{CMB}}$, respectively. In order to easily transition between analytic models, we leave the energy losses expressed in terms of pressure evolution rather than reducing $\rmn{d}\gamma/\rmn{d}t$ down to its time evolution as done in KDA97.
  Rearranging and taking the integral of both sides, we have
  \begin{equation}
  \frac{p_{\rmn{c}}(t)^{1/(3\Gamma_{\rmn{c}})}}{\gamma}-\frac{p_{\rmn{c}}(t_{\rmn{i}})^{1/(3\Gamma_{\rmn{c}})}}{\gamma_{\rmn{i}}}=f\left(t,\,t_{\rmn{i}}\right)\label{e:gamma evolution 2}
  \end{equation}
  where $f\left(t,\,t_{\rmn{i}}\right)$ is equivalent to KDA97's $a_{2}\left(t,\,t_{\rmn{i}}\right)$ and includes the radiative losses of the system.
  \begin{equation}
  f\left(t,\,t_{\rmn{i}}\right)=\int_{t_{\rmn{i}}}^t \frac{4\sigma_{\rmn{T}}}{3m_{\rmn{e}}c} p_{\rmn{c}}(t')^{1/(3\Gamma_{\rmn{c}})} \left[u_{\rmn{B}}(t') + u_{\rmn{CMB}} \right]\rmn{d}t' \label{e:f}
  \end{equation}
  An expression for the initial Lorentz factor of an electron at $t_{\rmn{i}}$ is found from equation (\ref{e:gamma evolution 2}), analogous to KDA97 equation (10).
  \begin{equation}
  \gamma_{\rmn{i}}\left(t,\,t_{\rmn{i}}\right)=\frac{\gamma p_{\rmn{c}}(t_{\rmn{i}})^{1/(3\Gamma_{\rmn{c}})}}{p_{\rmn{c}}(t)^{1/(3\Gamma_{\rmn{c}})}-f\left(t,\,t_{\rmn{i}}\right)\gamma}\label{e:initial gamma}
  \end{equation}
  Note that in these radiative equations, $p_{\rmn{c}}\left(t\right)$ and $V_{\rmn{c}}\left(t\right)$ refer to the full piecewise expressions, equations (\ref{e:full pressure}) and (\ref{e:full volume}). Thus there are cases near a phase boundary where $t_{\rmn{i}}$ is less than the transition time but $t$ is greater, such that terms with $t_{\rmn{i}}$ come from the previous phase's dynamical expressions. 
  
  The electron number density depends upon the relativistic electron energy density, $u_{\rmn{e}}(t_{\rmn{i}})$, at the time of injection, assuming an initial distribution $n(\gamma_{\rmn{i}},t_{\rmn{i}})\,\rmn{d}\gamma_{\rmn{i}}=n_{\rmn{0}}\gamma_{\rmn{i}}^{-p}\rmn{d}\gamma_{\rmn{i}}$. Here we use the kinetic energy of the electrons $(\gamma_{\rmn{i}} -1)m_{\rmn{e}} c^{2}$ to determine $n_{\rmn{0}}$, the initial number density of relativistic, radiating particles.
  \begin{equation}\label{e:no}
  n_{\rmn{0}}=\frac{u_{\rmn{e}}(t_{\rmn{i}})}{m_{\rmn{e}}c^{2}}\left[\intop_{\gamma_{\rmn{i},\,\rmn{min}}} ^{\gamma_{\rmn{i},\,\rmn{max}}} (\gamma_{\rmn{i}}'-1)\gamma_{\rmn{i}}'^{-p}\,\rmn{d}\gamma_{\rmn{i}}'\right]^{-1}
  \end{equation}
  Requiring the conservation of particles (i.e. no diffusion) as the cocoon volume expands then we have (cf. KDA97 equation 9)
  \begin{equation}\label{e:n(t)}
  n(t,\,t_{\rmn{i}})\,\rmn{d}\gamma=n_{\rmn{0}}\frac{\gamma_{\rmn{i}}\left(t,\,t_{\rmn{i}}\right)^{2-p}}{\gamma^{2}}\left(\frac{p_{\rmn{c}}(t)}{p_{\rmn{c}}(t_{\rmn{i}})}\right)^{4/(3\Gamma_{\rmn{c}})}\rmn{d}\gamma
  \end{equation}
  We assume that the cocoon magnetic field energy density is a fraction of the particle energy density such that $r=u_{\rmn{B}}/\left(u_{\rmn{e}}+u_{\rmn{T}}\right)$, where $u_{\rmn{T}}$ is the thermal particle energy density. $k'$ is defined as the ratio of the thermal particle energy density to the electron energy density. Then, the piecewise continuous expressions for electron and magnetic field energy densities are
  \begin{equation}\label{e:ue}
  u_{\rmn{e}}(t_{\rmn{i}})=\frac{p_{\rmn{c}}(t_{\rmn{i}})}{\left(\Gamma_{\rmn{c}}-1\right)\left(k'+1\right)\left(r+1\right)}
  \end{equation}
  \begin{equation}\label{e:ub}
  u_{\rmn{B}}(t_{\rmn{i}})=\frac{r p_{\rmn{c}}(t_{\rmn{i}})}{\left(\Gamma_{\rmn{c}}-1\right)\left(r+1\right)}
  \end{equation}
  To continue the approach of KDA97 we track radiative losses via small volume elements, $\delta V\left(t,\,t_{\rmn{i}}\right)$, labelled by their injection times, $t_{\rmn{i}}$. As the source grows, additional volume elements are injected and the older elements (corresponding to earlier injection times) suffer energy losses. For simplicity and generality, we derive the volume element expression from the total cocoon volume (equation \ref{e:full volume}) rather than assuming knowledge of how the hotspot pressure evolves relative to the cocoon pressure as done in KDA equation 14. In the self-similar model, the hotspot and cocoon pressure co-evolve but this is not expected in the young source model. Using equations (\ref{e:full length}) and (\ref{e:full volume}), we differentiate $V_{\rmn{c}}$ with respect to time and subtract the contribution by adiabatic expansion to get an expression for $\delta V\left(t_{\rmn{i}}\right)$ immediately after the element is injected into the cocoon. 
  \begin{equation}
  \delta V_{\rmn{c}}\left(t_{\rmn{i}}\right)=\dot{V}_{\rmn{c}}\left(t_{\rmn{i}}\right)\delta t_{\rmn{i}} - \delta V_{\rmn{c}}\left(t_{\rmn{i}}\right)_{\rmn{adiab}} \label{e:dVi 1}
  \end{equation}
  where
  \begin{equation}
  \dot{V}_{\rmn{c}}\left(t_{\rmn{i}}\right)=\frac{\rmn{d} V_{\rmn{c}}}{\rmn{d} D} \frac{\rmn{d}D\left(t_{\rmn{i}}\right)}{\rmn{d}t_{\rmn{i}}} \label{e:dV/dti}
  \end{equation}
  and $\delta V_{\rmn{c}}\left(t_{\rmn{i}}\right)_{\rmn{adiab}}$ comes from the requirement that $p_{\rmn{c}}V_{\rmn{c}}^{\Gamma_{\rmn{c}}}~=~\rmn{constant}$ for adiabatic expansion,
  \begin{equation}
  \delta V_{\rmn{c}}\left(t_{\rmn{i}}\right)_{\rmn{adiab}}=-\frac{1}{\Gamma_{\rmn{c}}}V\left(t_{\rmn{i}}\right)\frac{\dot{p}_{\rmn{c}}\left(t_{\rmn{i}}\right)}{p_{\rmn{c}}\left(t_{\rmn{i}}\right)}\delta t_{\rmn{i}} \label{e:dVadiab}
  \end{equation}
  where
  \begin{equation}
  \dot{p}_{\rmn{c}}\left(t_{\rmn{i}}\right)=\frac{\rmn{d} p_{\rmn{c}}}{\rmn{d} D} \frac{\rmn{d}D\left(t_{\rmn{i}}\right)}{\rmn{d}t_{\rmn{i}}} \label{e:dp/dti}
  \end{equation}
  Equations (\ref{e:dVi 1}) through (\ref{e:dp/dti}) express the initial volume element at time, $t_{\rmn{i}}$, just at the moment of injection into the cocoon from the hotspot. To propagate this to some time, $t$, later, we let this element adiabatically expand:
  \begin{equation}
  \delta V_{\rmn{c}}\left(t,\,t_{\rmn{i}}\right)=\delta V_{\rmn{c}}\left(t_{\rmn{i}}\right)\left(\frac{p_{\rmn{c}}\left(t_{\rmn{i}}\right)}{p_{\rmn{c}}\left(t\right)}\right)^{1/\Gamma_{\rmn{c}}}\delta t_{\rmn{i}} \label{e:dV}
  \end{equation}
  If this $\delta V_{\rmn{c}}\left(t,\,t_{\rmn{i}}\right)$ is integrated from $t_{\rmn{i}}$~=~$t_{\rmn{c}}$ to $t$, and from $t$~=~$t_{\rmn{c}}$ to $t$~=~$t_{\rmn{end}}$, then of course we recover the same total volume expression as given in equation (\ref{e:full volume}).

  We assume that for a given electron energy, radiation is emitted only at the critical frequency (which we set to the observing frequency, $\nu_{\rmn{obs}}$) and therefore we can use equation (\ref{e:critical frequency}) to express the current electron Lorentz factor in terms of the surrounding magnetic field strength and the critical frequency
  \begin{equation}\label{e:gamma}
  \gamma=\sqrt{\frac{2\piup\, m_{\rmn{e}} \nu_{\rmn{c}}}{e B\left(t\right)}}
  \end{equation}
  The magnetic field strength is related to the energy density, $B\left(t\right)~=~u_{\rmn{B}}\left(t\right)/\left(2 \mu_{\rmn{0}} \right)$, and $u_{\rmn{B}}$ expands with the particle fluid if $\Gamma_{\rmn{c}}=\Gamma_{\rmn{B}}$:
  \begin{equation}\label{e:magnetic energy density}
  u_{\rmn{B}}(t,\,t_{\rmn{i}})=u_{\rmn{B}}(t_{\rmn{i}}) \left(\frac{p_{\rmn{c}}(t)}{p_{\rmn{c}}(t_{\rmn{i}})}\right)=u_{\rmn{B}}(t)
  \end{equation}
  By substituting $u_{\rmn{B}}(t)$, $\gamma$, $n(t,\,t_{\rmn{i}})$, and $\delta V_{\rmn{c}}\left(t,\,t_{\rmn{i}}\right)$ into equation (\ref{e:dPower ti}) and integrating
  over $t_{\rmn{i}}$, the total power from a radio source (per frequency per solid angle) as a function of source age, $t$, is 
  \begin{equation}
  P_{\rmn{\nu}}=\intop_{t_{\rmn{min}}(t)}^{t}\frac{\sigma_{\rmn{T}}\, c\, u_{\rmn{B}}\left(t\right)}{6\piup\nu}\gamma^{3}n\left(t,\, t_{\rmn{i}}\right)\delta V_{\rmn{c}}\left(t,\, t_{\rmn{i}}\right)\rmn{d}t_{\rmn{i}}\label{e:luminosity integral w losses}
  \end{equation}
  This is equivalent to equation (16) in KDA97. The lower limit on the integral is the minimum injection time, $t_{\rmn{min}}(t)$, for which a volume element can still contribute to the total emission, and depends upon the strength of the magnetic field, the observing frequency, and energy loss history. Any element injected before $t_{\rmn{min}}(t)$ is not considered in the total luminosity integral. We define $t_{\rmn{min}}(t)$ by when the Lorentz factor, $\gamma_{\rmn{i}}$, of the injected electron at time $t_{\rmn{i}}$ must approach infinity in order to still be radiating at time $t$. By numerically solving equation \ref{e:initial gamma} for $t_{\rmn{i}}$ as $\gamma_{\rmn{i}}(t,\,t_{\rmn{i}}) \rightarrow \infty$, we determine $t_{\rmn{min}}(t)$ as a function of current age $t$. This minimum injection time is piecewise continuous across the phases and is plotted against source age in figure \ref{f:tmin plot}. The sharp increase in $t_{\rmn{min}}(t)$ at later source ages is due to the onset of strong inverse-Compton scattering off CMB photons.
  \begin{figure*}
    \centering{}
    \includegraphics[width=0.8\textwidth]{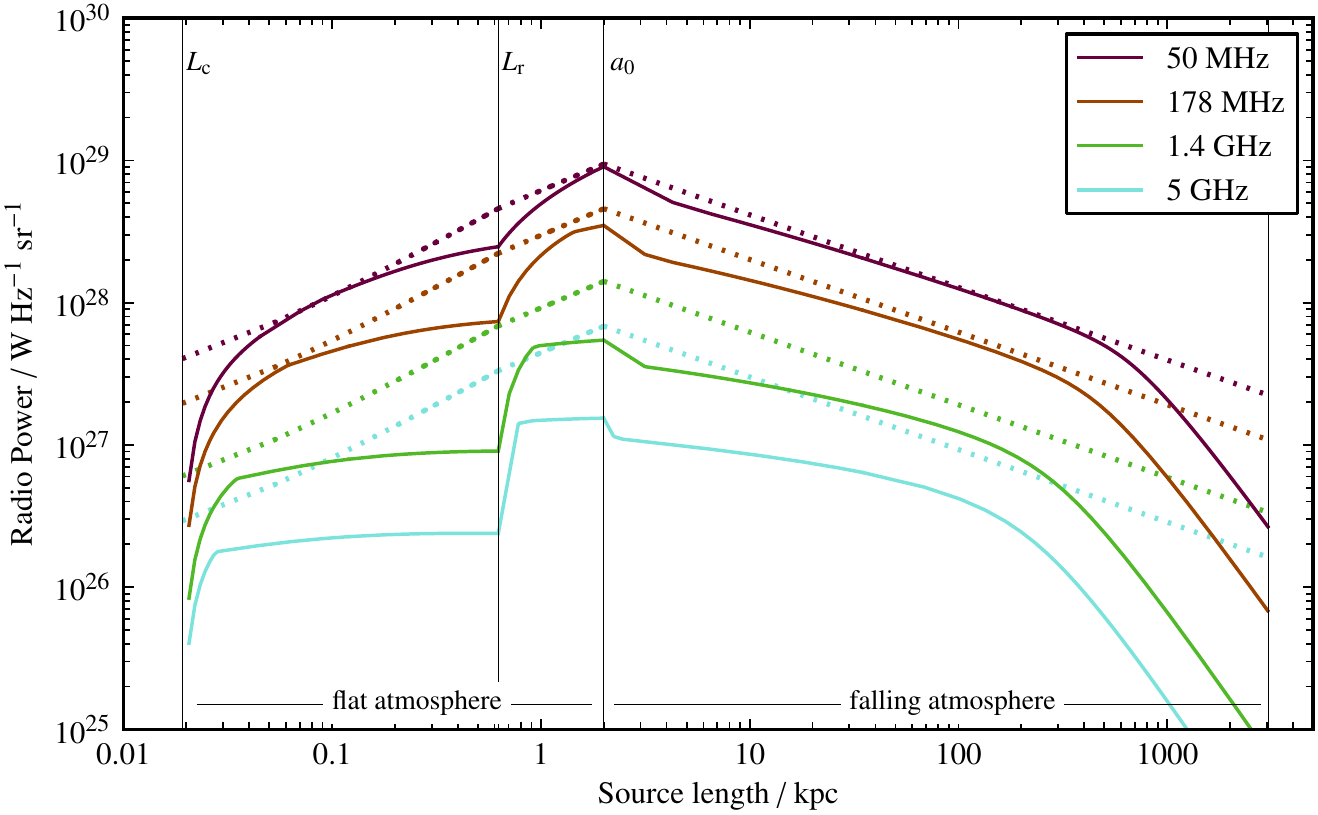}
    \caption{Radio source evolution through the $P{-}D$ diagram from first cocoon formation to the transition to a self-similar regime and intracluster scales, presented for observing frequencies, top to bottom, of $\nu_{\rmn{obs}}$ = 50 MHz (purple lines), 178 MHz (brown lines), 1.4 GHz (green lines), and 5 GHz (blue lines). The dotted lines show evolution without radiative ageing, the so-called `loss-less' tracks. The thick, solid tracks include radiative losses (equation \ref{e:luminosity integral w losses}) and show significant departure from the loss-less tracks for young sources, particularly at higher frequencies. Here we use $Q=1.3\times10^{39}$~W, $\theta=20^{\rmn{\circ}}$, and $z$ = 2. Tracks are computed from cocoon formation at $t_{\rmn{c}}$ until $t_{\rmn{end}}=10^{8}$~years. Other parameters as given in tables \ref{t:parameters} and \ref{t:derived parameters}. Vertical lines mark the size boundaries between the three phases discussed in section \ref{s:transition}.}
    \label{f:PD plot}    
  \end{figure*}
  \begin{figure}
    \centering{}
    \includegraphics[width=0.9\columnwidth]{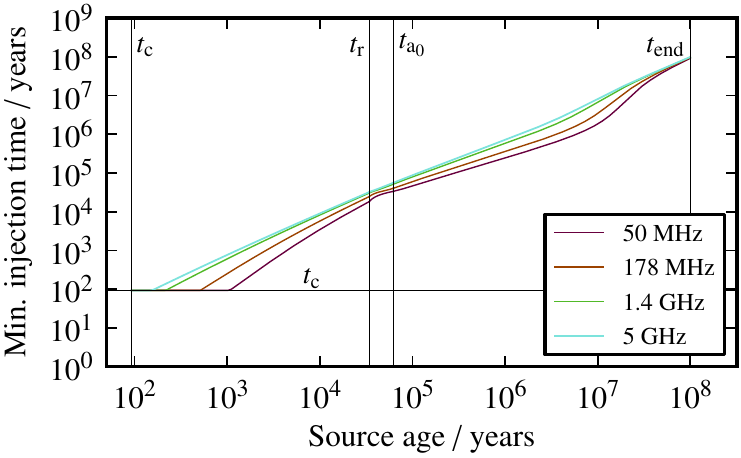}
    \caption{Minimum injection time, $t_{\rmn{min}}(t)$, from cocoon formation until end of life, presented for a range of observing frequencies, top to bottom, of $\nu_{\rmn{obs}}$ = 5 GHz (blue lines), 1.4 GHz (green lines), 178 MHz (brown lines), and 50 MHz (purple lines). See discussion of $t_{\rmn{min}}(t)$ in section \ref{s:energy losses}. Here we use $Q=1.3\times10^{39}$~W, $\theta=20^{\rmn{\circ}}$, and $z$ = 2. Other parameters as given in tables \ref{t:parameters} and \ref{t:derived parameters}. Vertical lines mark the size boundaries between the three phases.}
    \label{f:tmin plot}    
  \end{figure}
  The integral for total radio power is not analytically tractable but can be numerically evaluated from $t=t_{\rmn{c}}$ to $t_{\rmn{end}}$ to obtain the full path through the $P{-}D$ diagram for double radio sources. Figure \ref{f:PD plot} shows the evolutionary tracks for all three phases (solid lines), using equation (\ref{e:luminosity integral w losses}). Phase 1 emission is evaluated from $t=t_{\rmn{c}}$ to $t=t_{\rmn{r}}$ for $\beta=0$, using the dynamical expressions of A06. Phase 2 emission is evaluated from $t=t_{\rmn{r}}$ to $t=t_{\rmn{a}_{\rmn{0}}}$ for $\beta=0$ and with constants determined by the boundary conditions at $L_{\rmn{r}}$. Phase 3 is evaluated from $t=t_{\rmn{a}_{\rmn{0}}}$ to $t=t_{\rmn{end}} \sim 10^{8}$ years for $\beta>0$ and constants determined by the boundary conditions at $a_{\rmn{0}}$. For this plot we use the typical values for parameters listed in tables \ref{t:parameters} and \ref{t:derived parameters}.

  \begin{table}
  \caption{Parameters of the young source and self-similar models, unless otherwise specified.}
  \label{t:parameters}
  \begin{tabular}{ c c c}
    \hline
    Model parameter & Value & Reference \\
    \hline
    $Q$ & 1.3$\times10^{39}$~W~& 3 \\
    $\varv_{\rmn{j}}$ & $c$ & 1 \\
    $\theta$ & $20^{\rmn{\circ}}$ &  \\
    $\rho_{\rmn{0}}$ & 7.2$\times10^{-22}\, \rm{kg}\, \rm{m}^{-3}$ & 3 \\
    $a_{\rmn{0}}$ & $2\, \rm{kpc}$ & 2, 4 \\
    $\beta$ & 0, 1.9 & 3 \\
    $\overline{m}$ & $1.67\times10^{-27}\, \rm{kg}$ &  \\
    $T_{\rmn{0}}$ & $10^{7}$~K & 4 \\
    $p_{\rmn{0}}$ & ($k_{\rmn{B}} T_{\rmn{0}} \rho_{\rmn{0}})\,/\,\overline{m} = 6\times10^{-11}$~Pa &  \\
    $p$ & $2\alpha+1=2.14$ & 3 \\
    $\Gamma_{\rmn{c}}$ & $4/3$ & 4 \\
    $\Gamma_{\rmn{j}}$ & $4/3$ & 3, 4 \\
    $\Gamma_{\rmn{B}}$ & $4/3$ & 3, 4 \\
    $\Gamma_{\rmn{x}}$ & $5/3$ & 4 \\
    $\gamma_{\rmn{i},\,\rmn{min}}$ & $1$ & 3 \\
    $\gamma_{\rmn{i},\,\rmn{max}}$ & $\infty$ & 3 \\
    $\eta$ & 1 &  \\
    $k'$ & 0 & 3 \\
    $r$ & 3/4 & 4 \\
    $\chi$ & 1/3 & 1 \\
    $t_{\rmn{end}}$ & $10^8$ years & 5 \\
    $z$ & 2 & \\
    $u_{\rmn{CMB}}$ & $4\times10^{-14}(z+1)^{4}\,\rm{ J}\, \rm{m}^{-3}$ & 4 \\
    $\nu_{\rmn{obs}}$ & $50\, \rm{MHz}\,\left(178\, \rm{MHz}, 1.4\, \rm{GHz}, 5\, \rm{GHz}\right)$ &  \\
    \hline
  \end{tabular}
  \medskip
  \\
  (1) A06; (2) KA97; (3) KDA97; (4) \citet{Kaiser_2007b}; (5) \citet{Antognini_2012}
  \end{table}

  \begin{table}
  \caption{Model-derived parameters. Phase 1 refers to the young source model with a constant external environment profile, $\beta = 0$, out to reconfinement. Phase 2 refers to the self-similar model from reconfinement to core radius, $a_{\rmn{0}}$, with $\beta = 0$. Phase 3 refers to the self-similar model with an external environment falling with $\beta = 1.9$.}
  \label{t:derived parameters}
  \begin{tabular}{ c c c c}
    \hline
    Parameter & Phase 1 & Phase 2 & Phase 3 \\
    \hline
    $L_{1}$ &  24 pc & 24 pc & 24 pc\\
    $L_{\rmn{c}}$ &  19 pc & 19 pc & -\\
    $t_{\rmn{c}}$ & 93 years & 93 years & -\\
    $K_{1}$ & 1.2 & 1.2 & 1.2\\
    $a_{1}$ & 0.28 & 0.28 & 0.28\\
    $a_{2}$ & 0.74 & 0.74 & 0.74 \\
    $a_{3}$ & 0.14 & 0.14 & 0.14 \\
    $a_{4}$ & 0.87 & 0.87 & 0.87\\
    $\omega$ & 1.05 & 1.05 & 1.05\\
    $\lambda$ & 1.15 & 1.15 & 1.15 \\
    $L_{\rmn{r}}$ & 68 pc & 68 pc & -\\
    $t_{\rmn{r}}$ & 4$\times10^{4}$ years & 4$\times10^{4}$ years & -\\
    $t_{1}$ & - & 0.87 $t_{\rmn{r}}$ & - \\
    $c_{1}$ & - & 2.26 & 2.26 \\
    $c_{2}$ & - & 7.60 & 7.60\\
    $c_{3}$ & - & 0.76 & 0.76 \\
    $t_{\rmn{a}_{\rmn{0}}}$ & - & 6.7$\times10^{4}$ years & 6.7$\times10^{4}$ years\\
    $t_{2}$ & - & - & 0.23 $t_{\rmn{a}_{\rmn{0}}}$\\
    $b_{1}$ & - & - & 0.15\\
    $b_{3}$ & - & - & 0.76\\
    \hline
  \end{tabular}
  \end{table}

\section{Results}\label{s:results}
The final $P{-}D$ diagram showing our analytic evolutionary tracks for radio-loud sources from first cocoon formation to fully self-similar expansion is presented in figure \ref{f:PD plot}. The `loss-less' tracks are shown by dotted lines, rising in the dense core environment and falling in the more stratified environment. The solid lines include radiative losses. 
First we note that the radio luminosities across the phase boundaries match because of the underlying piecewise continuous physical parameters. At the boundaries, the decaying emission from previous phases as well as the newly-injected material of the current phase are included in the total luminosity.
The sharp features at the boundaries are caused by the fact that we are matching two different analytic models across three regimes and thus expect some discontinuity as a result. In reality, given a smooth reconfinement process and a smooth environmental profile, the evolution across the phase boundaries will also be smooth. 

  \subsection{Radiative ageing in young sources}
  On sub-kiloparsec scales, the evolutionary tracks of figure \ref{f:PD plot} (solid lines) show significant departure from the corresponding loss-less tracks (dotted lines), particularly at high frequencies. This is indicative of radiative losses becoming important for young sources just before they transition to full self-similar evolution and before they leave the core environment. Strong radiative losses have been predicted before -- see for instance the discussion and rough $P{-}D$ track in KB07 section 2. The work in this paper improves their model by including radiative losses for a more realistic, non-self-similar model of young source evolution.
  
  \subsubsection{Synchrotron break frequency evolution}
\begin{figure}
    \centering{}
    \includegraphics[width=0.9\columnwidth]{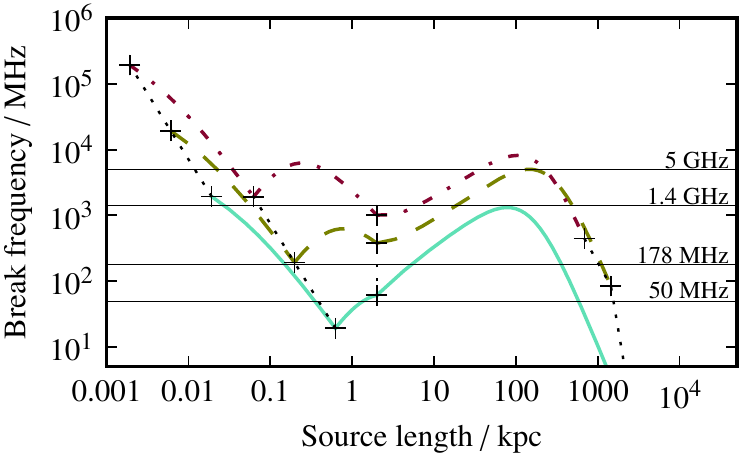}
    \caption{Break frequency as a function of source size. From top to bottom, purple dash-dot line: $Q=1.3\times10^{37}$~W, $z=0.2$, green dashed line: $Q=1.3\times10^{38}$~W, $z=0.5$, and blue solid line: $Q=1.3\times10^{39}$~W, $z=2$. Where the break frequency falls below the observing frequency, then radiative losses are significant. From left to right along each track, the black crosses joined by a dotted line mark $L_{\rmn{c}}$, $L_{\rmn{r}}$, $a_{0}$, and $L_{\rmn{end}}$, respectively.}
    \label{f:break freq plot}    
  \end{figure}

  We can further explore these radiative losses by examining the characteristic synchrotron break frequency, i.e the frequency above which the synchrotron spectrum falls away from a power-law due to preferential high-energy losses. Here we only consider losses due to synchrotron ageing and inverse-Compton scattering so that the break frequency is given by
  \begin{equation}
\nu_{\rmn{B}}\approx\frac{9 m_{\rmn{e}} e B\left(t\right)}{32\piup} \left[\frac{c}{\sigma_{\rmn{T}} t_{\rmn{lifetime}} (u_{\rmn{B}}(t)+u_{\rmn{CMB}})}\right]^{2} \label{e:break freq}
  \end{equation}
  and is plotted as a function of source size in figure \ref{f:break freq plot}.
  For this plot we simplistically assume that the magnetic field experienced by a given electron population is constant over its radiative lifetime, an assumption which is not made in the full radiative model. Thus the break frequency evolution plotted here is simply illustrative of the frequencies and length scales for which we expect strong ageing.

  Assuming an observing frequency of 50~MHz, and following the solid blue line of figure \ref{f:break freq plot} ($Q=1.3\times10^{39}$~W, $z=2$), we find that right before the transition to self-similar evolution at $L_{\rmn{r}}$ the break frequency drops below the observing frequency, implying significant radiative ageing due to synchrotron losses. This is caused by a `build-up' of the magnetic field in the slowly-expanding cocoon in the dense core environment. The injection of new relativistic electrons cannot compete with the severe synchrotron losses in the strong magnetic field. This disappears once the source is both reconfined and expanding in a more stratified environment such that the cocoon is growing fast enough to minimize radiative losses from the electrons. This agrees with the 50 MHz tracks of figure \ref{f:PD plot} where the gradient of the radiative track (solid purple line) is similar to the loss-less track (dotted purple line) implying that only adiabatic expansion losses are important. Continuing along the solid blue line of figure \ref{f:break freq plot}, additional radiative losses do not set in until inverse-Compton scattering effects at large scales. For higher observing frequencies, radiative losses are severe throughout the source evolution. Observing at 1.4 and 5~GHz for the typical parameters listed in table \ref{t:parameters} means observing above the break frequency at all times during the source lifetime. Large departures from the loss-less track are expected at these frequencies, which is confirmed by the higher frequency tracks in figure \ref{f:PD plot}. Break frequencies for smaller jet powers and lower redshifts are also plotted in figure \ref{f:break freq plot} for comparison and show much less ageing at low frequencies, as expected.

  \subsubsection{Spectral index evolution}
\begin{figure*}
    \centering{}
    \includegraphics[width=0.75\textwidth]{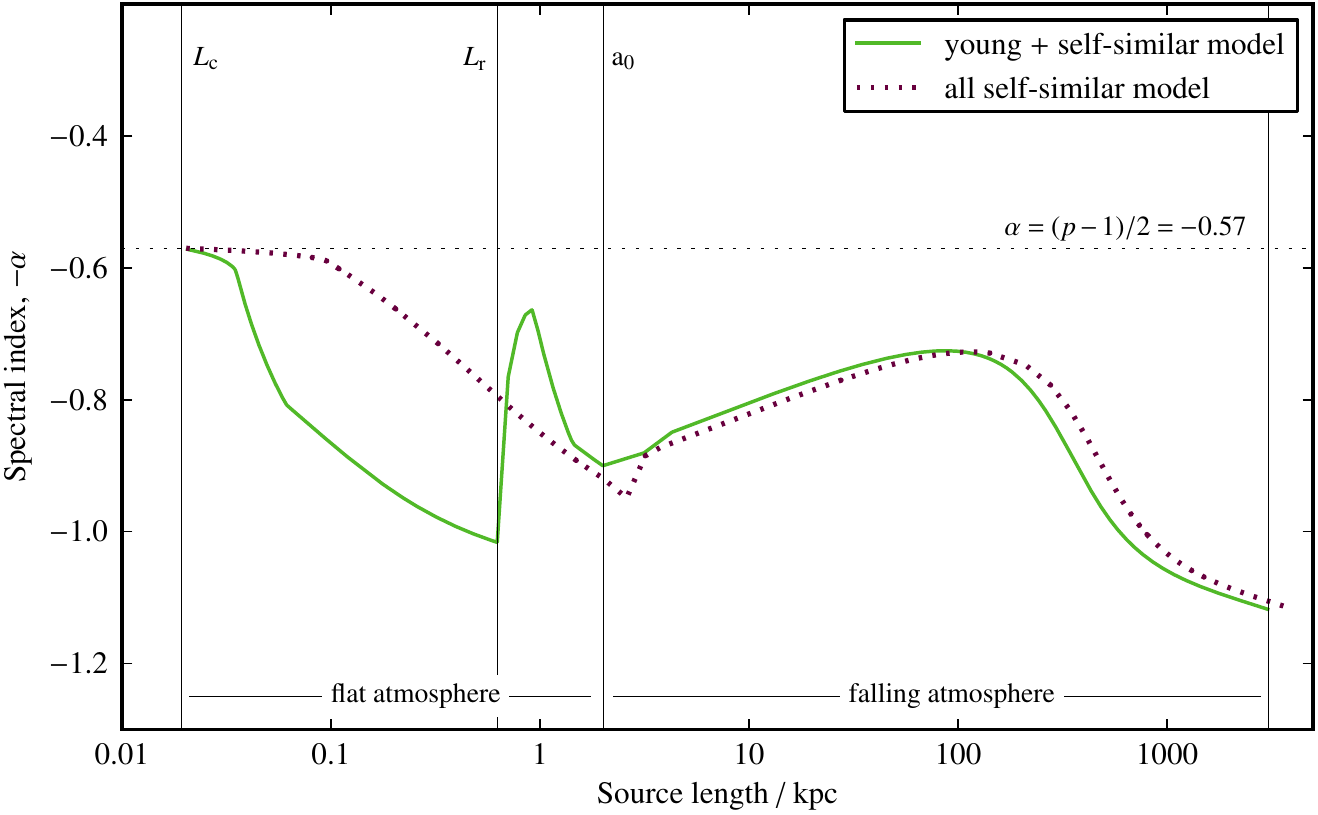}
    \caption{Cocoon spectral index, $-\alpha$, as it evolves with source size. $\alpha$ is measured between 178 MHz and 1.4 GHz, using the corresponding evolutionary tracks shown in figure \ref{f:PD plot} and assuming a power-law synchrotron spectrum. The solid green line uses the evolutionary model developed in this paper: young model for $D<L_{\rmn{r}}$ and self-similar KDA model for $D\geq L_{\rmn{r}}$. The dotted purple line assumes a simplistic extension of the self-similar KDA model for the entire source lifetime from $L_{\rmn{c}}$ onwards.}
    \label{f:spectral index}    
  \end{figure*}

  An estimate of how the cocoon spectral index evolves with source size can be derived using the $P{-}D$ tracks of figure \ref{f:PD plot} at $\nu_{\rmn{obs}}$~=~178~MHz and 1.4~GHz, assuming a power-law synchrotron spectrum, $F_{\rmn{\nu}}\propto \nu^{-\alpha}$. This spectral evolution, using the young model for $D<L_{\rmn{r}}$ and the self-similar KDA model for $D\geq L_{\rmn{r}}$, is shown in figure \ref{f:spectral index}, green solid line.
  Although this is a rough estimate of the overall cocoon spectral index, it does clearly predict a steepening of the cocoon spectra for young sources on sub-kiloparsec scales. Self-similar evolution in a falling atmosphere sees an initial flattening of $\alpha$ as radiative losses are minimal. The final spectral steepening occurs much later on scales of hundreds of kiloparsecs when inverse-Compton losses set in. At 100 kpc, we find an overall cocoon spectral index of $-7.2$, in good agreement with typical measured spectral indices at this scale.

  The scale of this spectral evolution matches observations of steep-spectrum CSOs and larger CSS sources, assuming that CSOs have sizes less than a kiloparsec and that CSS sources have kiloparsec sizes. For some source conditions, the rising spectral index at the beginning of phase 2 and phase 3 might translate to the growing awareness of flat-spectrum MSOs \citep{Augusto_2006}. We note that these scales change depending upon the particular jet power and environment density. But in general we suggest that young sources on sub-kiloparsec scales suffer from strong radiative losses and a steep spectral index. The process of reconfinement and the transition to a falling atmosphere both lead to a flattening of the spectral index, but IC losses at large scales steepen the total cocoon spectrum again. \citet{An_2012} discuss in detail an evolutionary path from CSO to MSO to large-scale double sources with a very similar progression from steep-spectrum to flat to steep again at large scales, summarized from the work of \citet{Kaiser_2007b}. Our young source model agrees well with these earlier discussions.
  
  For comparison, we can take a more simplistic approach to modelling young radio galaxies and extend the KDA self-similar model back to the cocoon formation scale, $L_{\rmn{c}}$. The corresponding evolutionary tracks, assuming self-similar expansion for all $D\geq L_{\rmn{c}}$, allow us to again compute the spectral index evolution (from 178~MHz and 1.4~GHz) for this fully self-similar model. This is also shown in figure \ref{f:spectral index}, purple dotted line. It predicts a flatter evolution of the spectral index on sub-kiloparsec scales compared to our young source model but, of course, a similar evolution on large scales. More investigation into this intriguing contrast between the self-similar and non-self-similar models at small scales is warranted, ideally with robust measurements of young source spectral indices that might differentiate between the models.

  \subsection{Effects of environment}\label{s:effects of environment}
  \begin{figure*}
    \centering{}
    \includegraphics[width=0.75\textwidth]{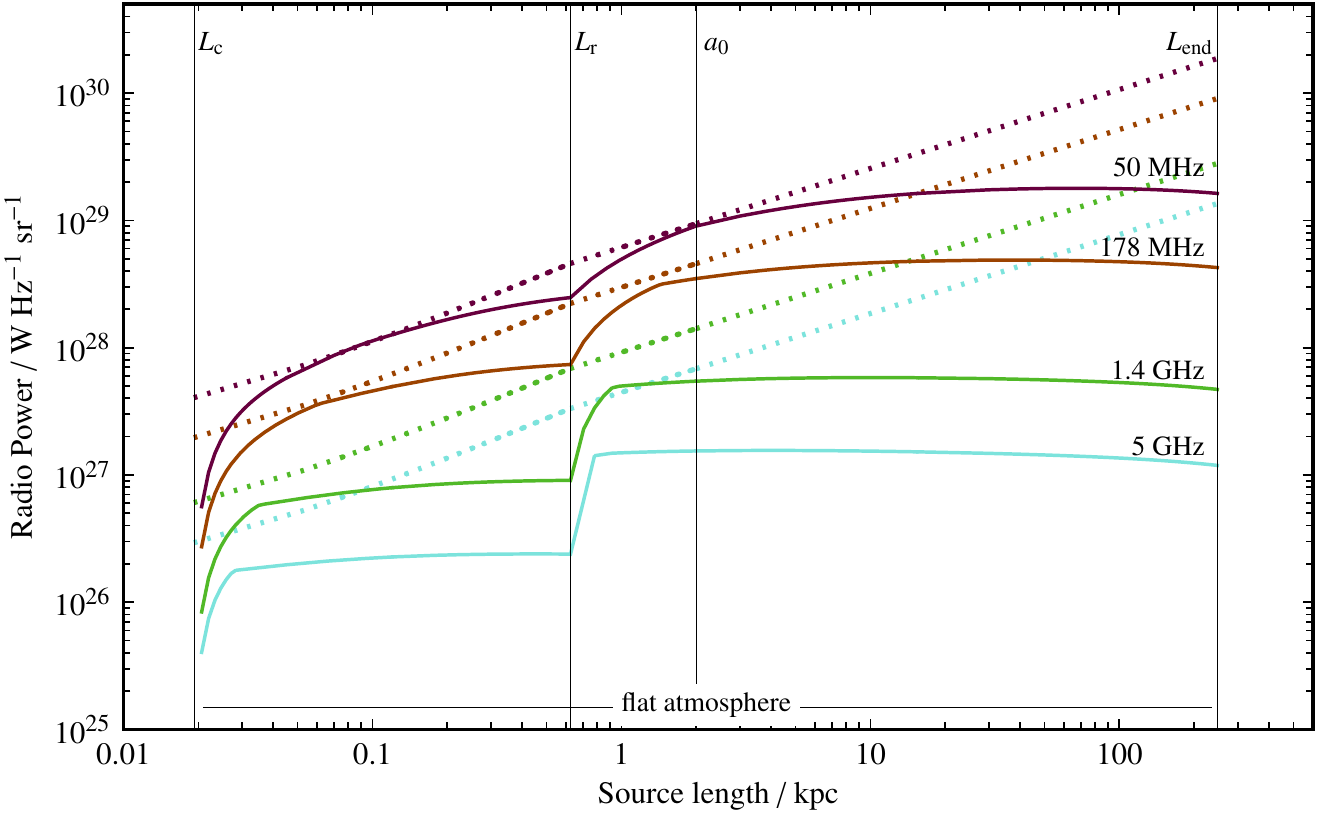}
    \caption{Radio source evolution through the $P{-}D$ diagram in a flat environment such that $\beta$ = 0 throughout, plotted for a range of observing frequencies, top to bottom: $\nu_{\rmn{obs}}~=~50$~MHz (purple lines), 178~MHz (brown lines), 1.4~GHz (green lines), and 5~GHZ (blue lines). The dotted lines show evolution without radiative ageing. The thick, solid tracks include radiative losses (equation \ref{e:luminosity integral w losses}) and show significant departure from the loss-less tracks in this flat environment. Here we use $Q=1.3\times10^{39}$~W, $\theta=20^{\rmn{\circ}}$, $z$ = 2, and $\rho_{0}=7.2\times10^{-22}$~kg~m$^{-3}$. Tracks are computed from cocoon formation at $t_{\rmn{c}}$ until $t_{\rmn{end}}=10^{8}$~years. As discussed in section \ref{s:effects of environment}, the corresponding maximum size, $L_{\rmn{end}}$, is dramatically shorter than the equivalent expansion in a falling atmosphere (figure \ref{f:PD plot}). Other parameters as given in tables \ref{t:parameters} and \ref{t:derived parameters}. Vertical lines mark the size boundaries between the three phases discussed in section \ref{s:transition}.}
    \label{f:PD flat plot}    
  \end{figure*}
  The external environment profile can have a dramatic effect on the evolution of radio galaxies. If we assume a constant density profile ($\beta=0$) for the entire lifetime of a radio source then the evolutionary track in phase 3 changes dramatically. In figure \ref{f:PD flat plot} we plot the `loss-less' tracks (dotted lines) and the radiative tracks allowing for spectral ageing (solid lines) for a range of observing frequencies. Strong energy losses due to the dense, flat environment are very apparent in phase 3, even at the low observing frequencies, and the luminosity as a function of source size is roughly constant. If we assume that the powering jet (using $Q=1.3\times10^{39}$~W, $\theta=20^{\rmn{\circ}}$) turns off after a typical lifetime of $t_{\rmn{end}}=10^{8}$~years, then in this flat environment the maximum source extent is only a few hundred kiloparsecs (from equation \ref{e:full length}) and IC losses have yet to set in. This can be compared to the equivalent evolutionary track in a flat then falling atmosphere as shown in figure \ref{f:PD plot}, where the radio source reaches sizes of 3000 kpc. This follows from the dependence of the source size on the external density (e.g. equation \ref{e:young size}).

\begin{figure*}
    \centering{}
    \includegraphics[width=0.75\textwidth]{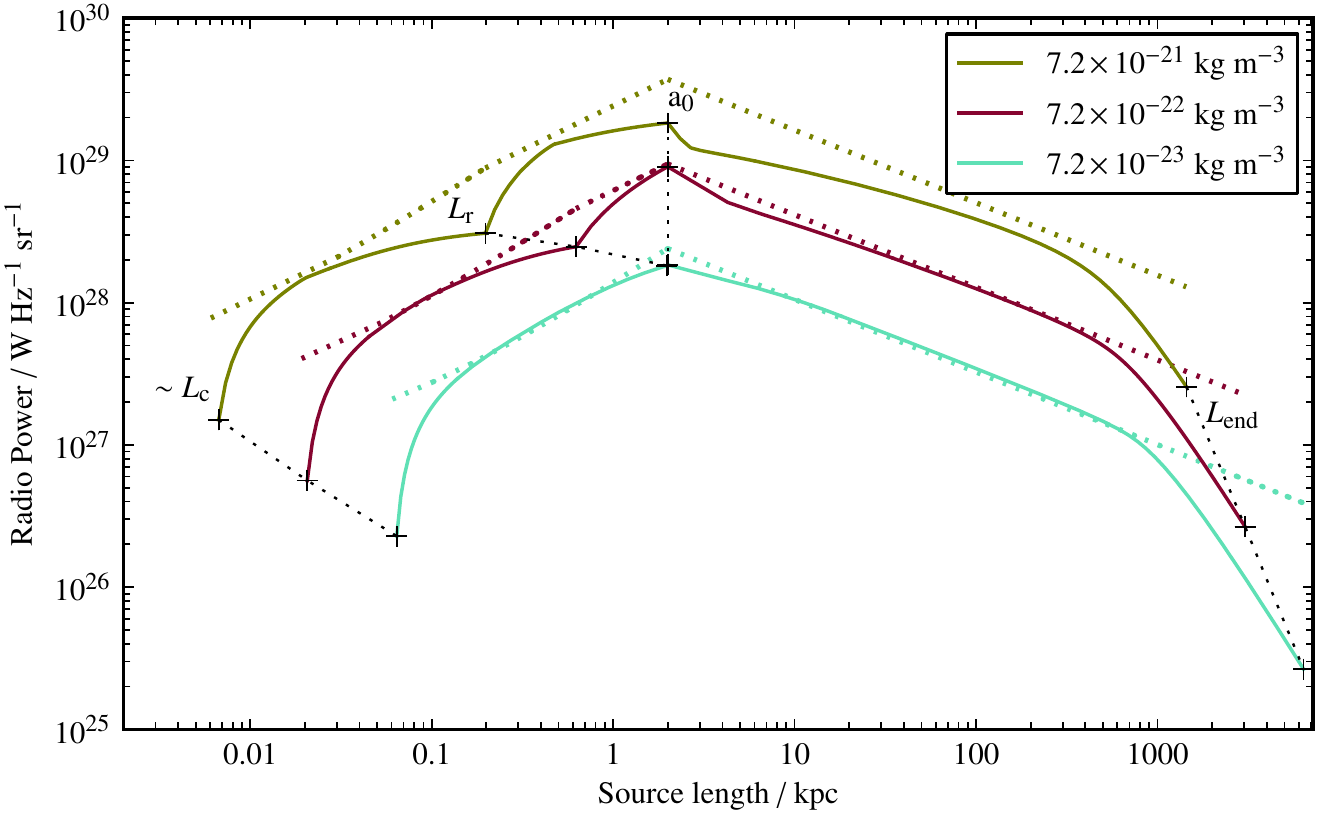}
    \caption{Radio source evolution in a range of external densities from cocoon formation until end of life at $t_{\rmn{end}}=10^{8}$ years. Dotted lines are the loss-less tracks and solid lines include radiative losses. The upper tracks are for $\rho_{0}=7.2\times10^{-21}$~kg~m$^{-3}$, the middle tracks are for $\rho_{0}=7.2\times10^{-22}$~kg~m$^{-3}$, and the lower tracks are for $\rho_{0}=7.2\times10^{-23}$~kg~m$^{-3}$. See discussion in section \ref{s:effects of environment}. Other parameters used: $\nu_{\rmn{obs}}=50$~MHz, $Q=1.3\times10^{39}$~W, $\theta=20^{\rmn{\circ}}$, $z$ = 2, and as listed in tables \ref{t:parameters} and \ref{t:derived parameters}. The phase transitions are marked for each track by the black crosses joined by a dotted line. $L_{\rmn{c}}$ is approximate since in this model, the cocoon luminosity starts at 0 at $D=L_{\rmn{c}}$ and very quickly reaches the luminosity shown at $\sim L_{\rmn{c}}$.}
    \label{f:PD density}    
  \end{figure*}

  As shown in figure \ref{f:PD density}, if the density of the external atmosphere is increased, then the radio source luminosity also increases, thanks to a higher cocoon pressure (cf. equations \ref{e:young pressure} and \ref{e:flat self-similar pressure with length}) and thus higher rates of synchrotron loss. The scales at which cocoon formation, jet reconfinement, and eventual end-of-life occur decrease as a direct consequence of the denser external environment. Figure \ref{f:PD density} plots three evolutionary tracks for $\rho_{0}=7.2\times10^{-23}$, $7.2\times10^{-22}$, and $7.2\times10^{-21}$~kg~m$^{-3}$.
  %

  \subsection{Effects of jet power and redshift}\label{s:jetpower redshift}
\begin{figure*}
    \centering{}
    \includegraphics[width=0.75\textwidth]{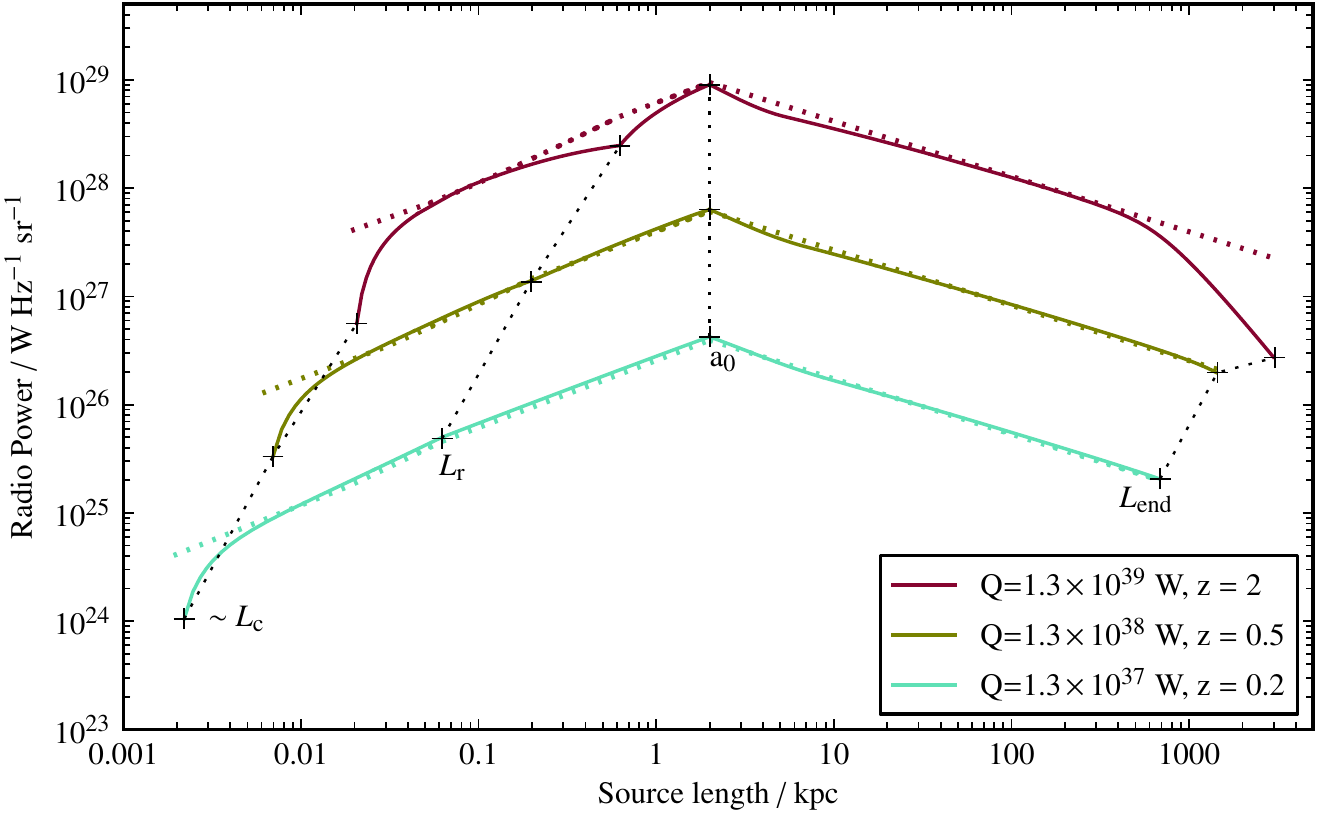}
    \caption{Evolutionary tracks for three radio-loud sources from cocoon formation until end of life at $t_{\rmn{end}}=10^{8}$ years. Dotted lines are the loss-less tracks and solid lines include radiative losses. The upper tracks are for $Q=1.3\times10^{39}$~W and $z$ = 2 (and are equivalent to the purple, middle tracks of figure \ref{f:PD density}). The middle tracks are for $Q=1.3\times10^{38}$~W and $z$ = 0.5. The lower tracks are for $Q=1.3\times10^{37}$~W and $z$ = 0.2. See discussion in section \ref{s:jetpower redshift}. Other parameters used: $\theta=20^{\rmn{\circ}}$, $\nu_{\rmn{obs}}=50$~MHz, and as listed in tables \ref{t:parameters} and \ref{t:derived parameters}. The phase transitions are marked for each track by the black crosses joined by a dotted line. $L_{\rmn{c}}$ is approximate since in this model, the cocoon luminosity starts at 0 at $D=L_{\rmn{c}}$ and very quickly reaches the luminosity shown at $\sim L_{\rmn{c}}$.}
    \label{f:PD jetpower redshift}    
  \end{figure*}

  The jet power, $Q$, greatly affects the shape of the overall evolutionary track and contributes to the amount of radiative ageing a radio source undergoes. In figure \ref{f:PD jetpower redshift} we plot radiative tracks for three different radio sources, varying the jet power and redshift (the latter determines the density of CMB photons and thus the strength of IC losses). A more powerful jet expands quickly and remains ballistic for longer, but suffers strong synchrotron losses in the dense core environment, even at low observing frequencies. A less powerful jet (i.e. $Q\sim 10^{37}$~W) forms a cocoon earlier and closely follows the loss-less track at low frequencies for the full evolution through the $P{-}D$ diagram. We present this plot as analogous to figure 1 in KDA97, albeit with a lower but overlapping range of jet powers and at a lower observing frequency of $\nu_{\rmn{obs}}=50$~MHz.
  \subsection{Hotspot contribution}\label{s:hotspot model}
\begin{figure}
    \centering{}
    \includegraphics[width=0.9\columnwidth]{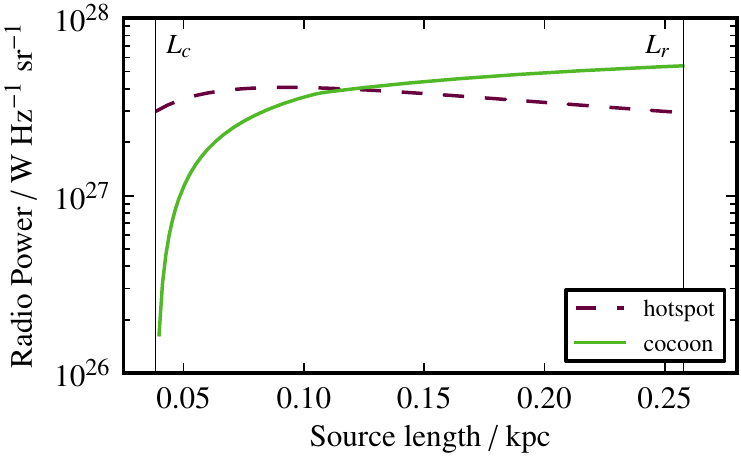}
    \caption{Simple hotspot model assuming no ageing plotted alongside the young cocoon model developed in section \ref{s:radiative model}. Parameters used: $Q=1.3\times10^{39}$~W, $\theta=10^{\rmn{\circ}}$, $c_{\rmn{h}}=0.5$, $\delta=0.5$, $\nu_{\rmn{obs}}=178$~MHz. Linear scale on x-axis.}
    \label{f:PD w hotspot}    
  \end{figure}

  The contribution from the hotspot is neglected in this paper under the assumption that the cocoon energetics quickly dominate over the hotspot for scales greater than $\sim L_{\rmn{c}}$. We can confirm this by constructing a simple model of hotspot luminosity, assuming minimum energy evolution without additional radiative losses, and to do so, an expression for the evolution of the hotspot pressure and volume before jet reconfinement is required. Hotspot pressure comes from the strong shock conditions at the jet terminus and is given by A06 equation (15):
  \begin{equation}
  p_{\rmn{h}}=\frac{2\alpha_{\rmn{x}}}{\Gamma_{\rmn{x}}+1} \frac{\rho_{\rmn{x}}\varv_{\rmn{j}}^{2}}{\left[\left(K_{1}D_{1}(t)/L_{1}\right)+1\right]^{2}} \label{e:hotspot pressure}
  \end{equation}
  where $\alpha_{\rmn{x}}$ is defined as $\alpha_{\rmn{x}}=\left[1+(\Gamma_{\rmn{x}}-1)^{2}/(4\Gamma_{\rmn{x}})\right]^{\Gamma_{\rmn{x}}/(\Gamma_{\rmn{x}}-1)}$.
  As discussed in A06 section 3.1, the hotspot volume can be approximated by a dimensional argument. The hotspot radius will scale with the unconfined jet radius, $r\propto \Omega^{1/2} D_{1}(t)$, and the hotspot length along the jet axis will likely scale with some combination of $L_{1}$ and $D_{1}(t)$, such that $V_{\rmn{h}}\,=\,c_{\rmn{h}}\left[\Omega^{1/2} D_{1}(t)\right]^{2+\delta}L_{1}^{1-\delta} $, where $c_{\rmn{h}}$ < 1 and $0 \leq \delta \leq 1$. With these expressions the hotspot luminosity will evolve according to equation (\ref{e:min energy luminosity}). In figure \ref{f:PD w hotspot} we plot this simple hotspot model alongside the phase 1 young source model (equation \ref{e:luminosity integral w losses}) on the $P{-}D$ diagram. Here the length axis is a linear scale to emphasis how soon the hotspot contribution to the total radio luminosity ceases to be significant. We assume the hotspot material is relativistic ($\Gamma_{\rmn{h}}=4/3$) and comprised only of light particles (no proton content) such that $\eta=1$.
  As the cocoon evolves beyond $L_{\rmn{c}}$, the hotspot luminosity soon ceases to be dominant even with the conservative parameters we use in figure \ref{f:PD w hotspot}. Thus in the model of young source evolution, we ignore the energetics of the hotspot. This is further justified in the self-similar phase due to the much larger relative size of the cocoon compared to the hotspot. Analytic models of radio sources that push to younger ages before cocoon formation will of course need to include the hotspot (and jet) contribution to the total radio emission.

\section{Discussion}\label{s:discussion}
In this section we consider the applicability of the model presented in this work, and in particular the physical scenarios where the young source model might break down.

The dynamical model developed by A06 assumes that a young source expands into a flat atmosphere and that the jet undergoes reconfinement at $L_{\rmn{r}}$ before transitioning beyond the flat galaxy core. This model does not encompass regimes where reconfinement might occur on scales beyond the core radius in strongly declining atmospheres. Thus we should consider two potential issues: does reconfinement occur within a flat core and if not, is the density gradient small enough to be neglected in the model?

To address both issues, we must consider the scaling of $L_{\rmn{r}}$. This is comparable to considering the scaling of the characteristic length scale, $L_{\rmn{1}}$, since from equation \ref{e:reconfinement}, $L_{\rmn{r}}\propto L_{\rmn{1}}$. $L_{\rmn{1}}~\propto~Q^{1/2}\,\rho_{0}^{-1/2}\,\varv_{\rmn{j}}^{-3/2}$ so we can immediately see that for radio sources with large jet powers and low density environments, $L_{\rmn{1}}$, and thus $L_{\rmn{r}}$, becomes large and the assumption that $L_{\rmn{r}}$ is less than the core radius may no longer hold.

The reconfinement length scale also depends upon $\Omega$ ($= \piup \theta^{2}$) and our choice of $\alpha$. Increasing the opening angle increases $L_{\rmn{r}}$ so that a powerful jet with a large opening angle might also reconfine beyond the core radius. The choice of $\chi=1/3$ comes from the approximate argument that several reconfinement shock structures of length $2 z_{1}$ must fit into the total length of the jet before the jet can be considered to be fully reconfined and the source expanding self-similarly. This agrees well with simulations of FRII sources by \citet{Komissarov_1998} which find that for jets with $\theta=20^{\circ}$, which we consider throughout this paper, self-similar evolution begins very soon after reconfinement (see their figure 3). But they also find that the start of self-similar evolution depends upon the jet opening angle where smaller jet angles seem to require more jet structures, and thus smaller values of $\alpha$ before self-similar evolution begins. A decrease in $\alpha$ increases $L_{\rmn{r}}$. Determining when full self-similar evolution begins is complex (see also \citealt{Carvalho_2002}) and due to its uncertainty, we leave $\alpha$ as a parameter of the model.

Typical reconfinement scales are represented in figures \ref{f:PD density} and \ref{f:PD jetpower redshift} for a range of core densities and jet powers, respectively, and are consistently less than a few kiloparsecs and thus likely to be within a flat galaxy core radius. However, recent hydrodynamical simulations of galactic environments by \citet{Barai_2013} predict significant density gradients even at kiloparsec scales (the resolution limit of their simulation). This brings us to the second issue. If there is a real density gradient on scales relevant to our young source model then the gradient must be less than the characteristic gradient of the model in order for the model to still be applicable. In other words, this model is valid for external mediums where $\rmn{d}\rho_{\rmn{x}}/\rmn{d}r < \rho_{0}/L_{\rmn{1}}$.

The young source model has been motivated by the need to consider mechanical AGN feedback within galaxies, and low-power radio sources, which comprise the majority of radio-loud sources in the universe, are expected to contribute the most to this feedback process. Given the scaling of $L_{\rmn{r}}$, it is these types of sources which should be well-modelled by the dynamical and radiative expressions presented in A06 and this paper.

\section{Summary}\label{s:summary}
A model for the radiative evolution of young radio sources on sub-kiloparsec scales has been developed as an extension of the earlier non-self-similar dynamical model of A06. This radiative evolution treats energy loss mechanisms within the cocoon in the same manner as the KDA97 model does for larger double sources. The complete description of radio source evolution now extends from the moment of cocoon formation on parsec scales to when the cocoon pressure reconfines the conical jet (the so-called phase 1); through the subsequent transition to self-similar expansion in flat and falling atmospheres (phase 2 and 3); and the eventual end of life around $10^{8}$~years on megaparsec scales.

By fully accounting for energy loss mechanisms, we find that synchrotron ageing is significant for the young radio sources, even at low frequencies. This is due to the dense environment on sub-kiloparsec scales and the declining forward momentum of the jet which leads to strong magnetic fields within the cocoon. Synchrotron ageing becomes much less significant once the jet is reconfined by the cocoon and the source transitions to a declining density environment. In phase 3, the forward momentum of the jet is increased and the newly-injected particles are able to compensate for synchrotron losses. Thus the self-similar track follows a `loss-less' path where only adiabatic expansion is important. As discussed in the \citet{Kaiser_1997b} model, at large-scales, the energy density of the cocoon electrons becomes comparable to the CMB photon energy density and inverse-Compton scattering triggers a sharp drop in radio luminosity on megaparsec scales from the central AGN.

We find that the dynamical model of A06, coupled with full radiative loss treatment, predicts a greater amount of ageing compared to a more simplistic extension of a self-similar model back to cocoon formation scales. This implies a steeper cocoon spectral index at small scales for the non-self-similar model and the potential for observational data to differentiate between radiative models of young sources.

This new model describing the radiative evolution of young radio sources will form an important framework on which to base the analysis of compact symmetric objects (CSOs) and should aid studies of CSO and large-scale FRII source counts as well as the broader picture of radio source evolution from galaxy core to intracluster medium. 


\section*{Acknowledgements}
T.M. thanks the Marshall Aid Commemoration Commission and St John's College, Cambridge for PhD research funding. We are most grateful to the referee for the detailed and helpful report. We would also like to thank Dave Green and Richard Butler for their comments on the manuscript. This paper makes use of the CubeHelix colour scheme of \citet{Green_2011}.

\setlength{\labelwidth}{0pt} 
\footnotesize{
\bibliographystyle{mn2e} 
\setlength{\bibsep}{0pt}
\renewcommand{\bibname}{References} 
\bibliography{research}
%



\bsp

\label{lastpage}

\end{document}